\documentclass[aps,pre,reprint,amsmath,amssymb,superscriptaddress,nofootinbib]{revtex4-2}

\usepackage{graphicx}
\usepackage{bm}
\usepackage{hyperref}
\usepackage{xcolor}
\newcommand{\Vxx}{\ensuremath{V_{xx}}}
\newcommand{\Vxy}{\ensuremath{V_{xy}}}
\newcommand{\Ixx}{\ensuremath{I_{xx}}}

\newcommand{\dVxy}{\ensuremath{\Delta V_{xy}}}

\newcommand{\neff}{\ensuremath{n_{\mathrm{eff}}}}
\newcommand{\AF}{\ensuremath{A_{F}}}
\newcommand{\kF}{\ensuremath{k_{F}}}

\newcommand{\degC}{\ensuremath{^\circ\mathrm{C}}}

\newcommand{\uL}{\ensuremath{\mu\mathrm{L}}}

\usepackage[T1]{fontenc}

\usepackage{float}

\begin{document}

\title{A Polarization Hall Effect in Hydrated DNA}
\author{Mariusz Pietruszka}
\email{mariusz.pietruszka@us.edu.pl}
\affiliation{University of Silesia, Faculty of Natural Sciences, Institute of Biology, Biotechnology, and Environmental Protection, 40-032 Katowice, Poland}
\date{\today}

\begin{abstract}
We report magnetic-field- and temperature-dependent transverse-voltage phenomena in hydrated genomic DNA. The sample is confined beneath a glass coverslip in a quasi-two-dimensional geometry and subjected to perpendicular magnetic fields, revealing sharp thresholds, staircase-like voltage plateaus, and oscillatory structures approximately periodic in inverse magnetic field. Fixed-field cooling reveals longitudinal reorganizations near 20.6 and 12.0~$^\circ$C, while a separate experiment shows large-amplitude transverse oscillations below approximately 12~$^\circ$C. Water controls and buffer-free DNA measurements support an association between the structured response and the hydrated DNA system. We propose a polarization-based interpretation of the transverse response, termed a ``polarization Hall effect,'' and discuss the observed regimes within a phenomenological framework of nonlinear dynamics and metastable configurations. These findings identify hydrated DNA as a soft-matter system exhibiting structured electrical responses jointly regulated by magnetic field and temperature.
\end{abstract}


\maketitle

\section{Introduction}

Hydrated DNA forms a structured yet dynamically fluctuating hydrogen-bond network whose collective electrodynamic properties remain only partially understood. At the molecular scale, proton transfer along hydrogen bonds, hydration-layer ordering, and long-range dipolar interactions have been studied extensively, with theoretical models predicting cooperative proton dynamics and correlated vibrational modes in soft biological assemblies~\cite{Frohlich1968,Frohlich1988,Davydov1979,HammesSchiffer2006,DonadioGalli2025}. These microscopic ideas are complemented by macroscopic frameworks developed to describe cooperative or synergistic organization in biological systems~\cite{Reimers2009,Vasconcellos2012}, as well as by broader discussions of collective organization in biological matter~\cite{AlKhaliliMcFadden2015}.

A central open question is whether such collective behavior can persist at mesoscopic scales under ambient conditions in the form of reproducible, field-organized polarization responses, given the well-known constraints imposed by thermal fluctuations and environmental noise in biological environments~\cite{Tegmark2000}. In an electrolyte with ideally blocking electrodes, double-layer charging screens the bulk DC field at equilibrium~\cite{Bazant2004}. Here the applied voltage sets the electrical boundary condition, while magnetic field and temperature are varied to probe the transverse potential. The experimental question is whether the time dependence of that potential, including the observed thresholds and oscillations, is consistent with a collective polarization response. Prior analyses of cooperative protonic and dipolar dynamics in biological matter, together with computational studies of hydrogen-bonded water~\cite{DonadioGalli2025}, suggest that a structured hydration shell could, in principle, support correlated dipolar fluctuations. Direct macroscopic evidence for such field-induced collective behavior, however, has remained limited.

Over the past several years, magnetic-field experiments on hydrated DNA have revealed a reproducible pattern of sudden voltage thresholds, staircase-like plateaus, and oscillatory features at or near room temperature~\cite{PietruszkaMarzec2024,Pietruszka2025a,Pietruszka2025b}. These observations show a formal resemblance to $1/B$-periodic structures familiar from magnetotransport studies~\cite{Shoenberg1984,Tsui1982,VonKlitzing1980}, while occurring in a biological soft-matter environment where conventional electronic transport mechanisms are not expected to dominate. Selected water and metallic controls show responses different from those recorded in hydrated DNA. These comparisons support an association between the observed voltage structures and the DNA-containing sample, motivating investigation of the contributions of hydration, molecular organization, and sample--electrode interfaces.

The physical interpretation of these features has therefore required reassessment. The threshold and staircase-like transverse-voltage response reported in Physica B were initially interpreted in terms of protonic charge transport and Landau-level-like quantization~\cite{Pietruszka2025a}. Subsequent temperature-dependent measurements showed a strong suppression of the longitudinal current while transverse voltage and photovoltage remained finite and structured, motivating the ohmic--polarization crossover interpretation developed in BioSystems~\cite{Pietruszka2026BioSystems}. The present work retains the reported voltage structures and extends their interpretation to include polarization dynamics. The earlier Physica B measurement at $21.9\,^\circ\mathrm{C}$ lies in a temperature range where conductive and polarization contributions may coexist; the cooling results therefore do not by themselves determine the mechanism of that earlier response. This interpretation allows conductive and polarization contributions to coexist; their relative importance may vary with temperature, magnetic field, and sample conditions. Temperature-dependent bistable switching has also been reported in hydrated genomic DNA under physiological ionic conditions~\cite{Pietruszka2026Bistable}.

Importantly, transverse responses carried by neutral excitations have become a recognized theme in modern condensed-matter physics. Experimental and theoretical studies have demonstrated that phonons, magnons, and excitons can generate Hall-like transverse signals through mechanisms involving chirality, broken symmetry, or geometric phase effects~\cite{Strohm2005,Onose2010,Mak2018,Saito2019,Li2023,ZhangGaoChen2024,Jin2025,Owerre2016}. These developments provide a conceptual context in which transverse responses in strongly polarizable, chiral soft-matter systems may be considered without requiring itinerant charge carriers. The intrinsic chirality of DNA and the polarizable hydrogen-bond network of its hydration environment motivate consideration of collective dipolar dynamics as a possible mechanism for the measured transverse response. Establishing this connection quantitatively requires a description of magnetic coupling and its relation to the detected voltage.

From this perspective, hydrated DNA can be investigated as a polarizable soft-matter system in a quasi-two-dimensional sample geometry, with magnetic field and temperature serving as control parameters for its electrical response. The possibility of collective organization in such dipolar ensembles is consistent with early predictions for driven biological systems~\cite{Frohlich1968,Reimers2009} and with general principles of mode selection and synchronization studied in many driven physical systems, including condensed-matter and nonlinear dynamical media~\cite{CrossHohenberg1993,Anderson1995,Ketterle1999}. Although the DNA--water interface operates in a regime very different from crystalline solids or ultracold gases, the underlying concept—the emergence of organized collective dynamics from coupled degrees of freedom—remains relevant.

In the present work, the DNA--water sample is confined beneath a glass coverslip, forming a hydrated layer whose lateral dimensions substantially exceed its thickness. We use the term quasi-two-dimensional to describe this sample geometry and the arrangement of longitudinal and transverse voltage contacts in the sample plane. The molecular hydration environment remains locally three-dimensional, while the planar measurement geometry provides a practical framework for investigating the field-dependent electrical response. Within this setting, polarization dynamics offer a possible interpretation of the observed oscillatory structures and voltage plateaus.

Here, we employ a combined magnetic-field and temperature-control strategy to map the regimes in which the transverse response emerges in hydrated DNA. By treating magnetic field and temperature as independent control parameters and by monitoring longitudinal and transverse signals acquired within the same measurement run, we resolve transitions between fluctuating regimes, oscillatory behavior, plateau-like states, and low-temperature large-amplitude modes. Together, these measurements reveal structured transverse-voltage responses whose thresholds, oscillatory patterns, and plateau-like features depend on magnetic field and temperature, providing an experimental basis for investigating nonlinear dynamics in hydrated DNA.

We use the term ``polarization Hall effect'' to describe the field-dependent transverse electrical response within the proposed polarization framework. The Hall analogy refers to the measurement geometry, with the transverse voltage recorded perpendicular to the longitudinal bias and the magnetic field applied normal to the sample plane. The physical hypothesis is that field-dependent reorganization of the hydrated DNA system modifies its polarization response, producing the observed voltage structures. This interpretation motivates the phenomenological model developed below.

We describe the staircase-like response in terms of discrete voltage levels and plateau spacing. Where the term ``quantized'' is retained in discussing earlier work, it denotes the operational identification of distinct macroscopic response levels. Their physical origin is examined within a phenomenological framework of metastability and field-dependent switching.

Building on the reported ohmic--polarization crossover~\cite{Pietruszka2026BioSystems}, we investigate how magnetic field and temperature shape the transverse-voltage response of hydrated DNA. We first present the measured thresholds, oscillatory regimes, and plateau-like structures, followed by control comparisons and relationships between detection channels. We then discuss a minimal dynamical model in which metastable configurations and switching provide a possible explanation of the observed response. This organization connects the experimental findings with a testable phenomenological interpretation.

\section{Results}

\subsection*{Field- and Temperature-Dependent Transverse Response in Hydrated DNA}

The measured observable is the transverse voltage $V_{xy}$ recorded across the hydrated DNA sample. We first characterize its dependence on magnetic field and temperature, then examine polarization dynamics as a possible physical interpretation of the observed structures.

The transverse-voltage response displays oscillatory structures and distinct plateau-like levels. A possible physical framework involves proton-mediated dipolar rearrangements within the hydrogen-bond network of the DNA--water environment. In this picture, collective reorientation contributes to a time-dependent polarization $P(t)$ and an associated polarization current density $J_{\mathrm{pol}}=\mathrm{d}P/\mathrm{d}t$. The measured voltage depends on how these dynamics couple to the sample geometry, electrodes, and measurement circuit.

Across the magnetic-field measurements, we identify three characteristic features: (i) oscillatory structures approximately periodic in $1/B$ in the transverse and longitudinal channels; (ii) a sharp voltage threshold separating distinct response regimes; and (iii) staircase-like plateaus in $V_{xy}$. Their occurrence and characteristic scales are examined below for the corresponding sample concentrations, temperatures, and sweep protocols.

Analysis of oscillatory periodicity and plateau spacing yields comparable effective density scales, with representative estimates near $n_{\mathrm{eff}}\approx4.7\times10^{14}\,\mathrm{m}^{-2}$ within the adopted phenomenological parameterization. This agreement provides a quantitative comparison between the two types of voltage structure. In selected measurements, a threshold near $0.35$~T separates dense oscillatory features from a response with broader lobes, suggesting a reorganization of the underlying dynamics. The water-control trace provides a reference for comparison with the structured DNA response.

We use inverse-field periodicity, plateau spacing, and tensor-based representations to characterize and compare the measured responses. Parameters derived through transport-inspired relations are treated as effective scales within the proposed polarization framework. Their interpretation depends on the assumptions connecting the measured voltages to the underlying dynamics, which are specified alongside the corresponding analysis.

Thresholds, plateau-like structures, and oscillatory features recur across the datasets presented here. We compare their characteristic scales with reference to the sample concentration, temperature, and acquisition protocol of each measurement.

\subsubsection*{Field-induced polarization response in DNA--water matrices}

\begin{figure}[t]
  \centering
  \includegraphics[width=\linewidth]{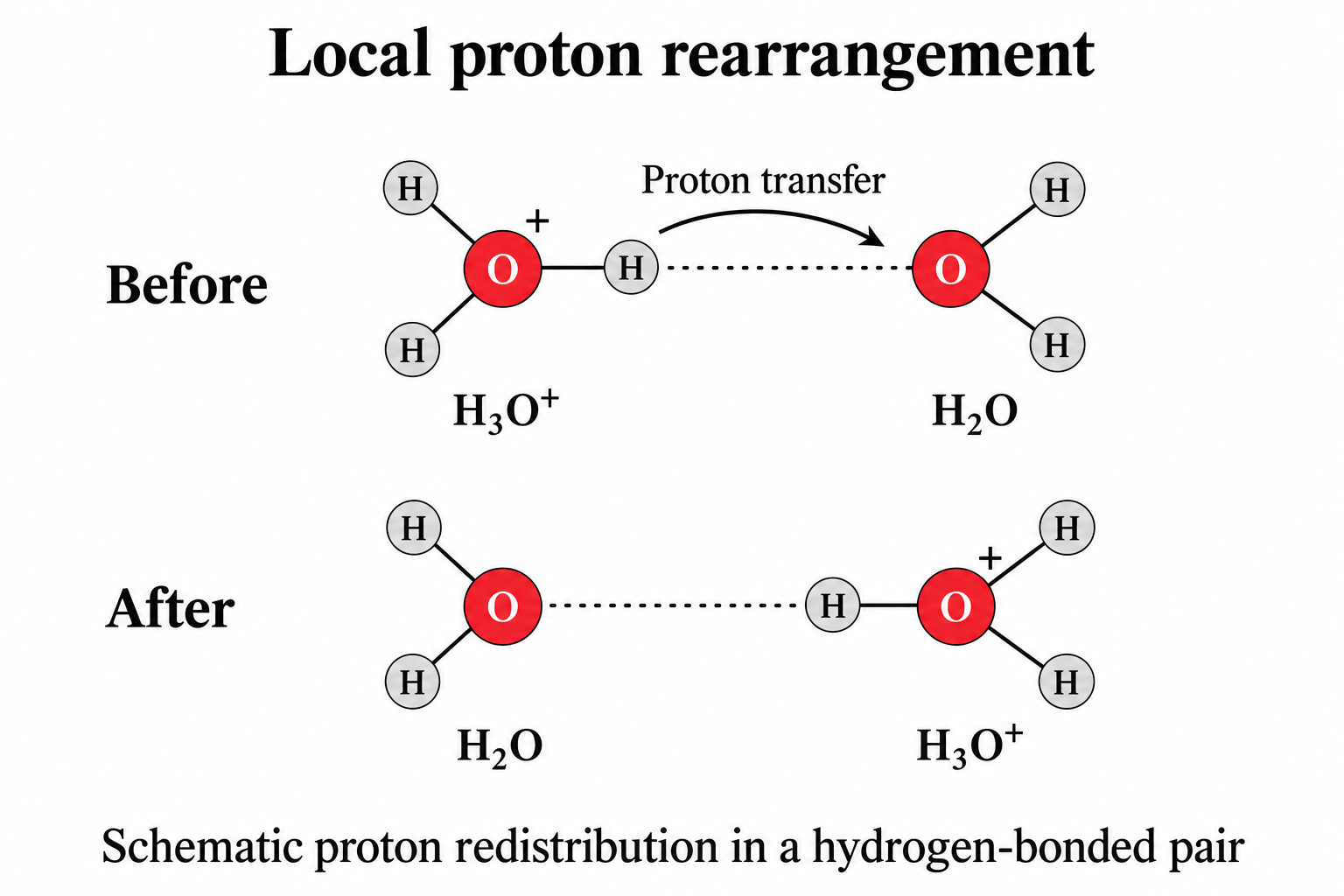}
\caption{Schematic proton transfer within a hydrogen-bonded hydronium--water pair. Before transfer, the donor is H$_3$O$^+$ and the acceptor is H$_2$O; after transfer, their roles are exchanged. Solid lines denote covalent O--H bonds, the dotted line denotes a hydrogen bond, and the arrow indicates the direction of proton transfer. The process redistributes the local positive charge while conserving the total charge and number of atoms. Such local rearrangements are considered as possible microscopic contributions to polarization dynamics in the hydrated DNA environment. The schematic was prepared with the assistance of generative AI.}
  \label{fig:fig1}
\end{figure}

We measured the transverse voltage \Vxy\ across a hydrated DNA sample confined in a quasi-two-dimensional geometry under a perpendicular magnetic field. The measurements reveal thresholds, oscillatory structures, and plateau-like levels under the conditions specified for each dataset. Figure~\ref{fig:fig1} illustrates a proposed microscopic picture in which transient proton transfer along hydrogen bonds produces local dipolar rearrangements. Coupling among these rearrangements could contribute to collective polarization dynamics and thereby influence the measured transverse response. This picture provides a physical hypothesis for interpreting the voltage structures and motivates the phenomenological model presented below.

\subsubsection*{Temperature-driven reorganization of longitudinal and transverse responses}

\begin{figure}[t]
  \centering
  \includegraphics[width=\linewidth]{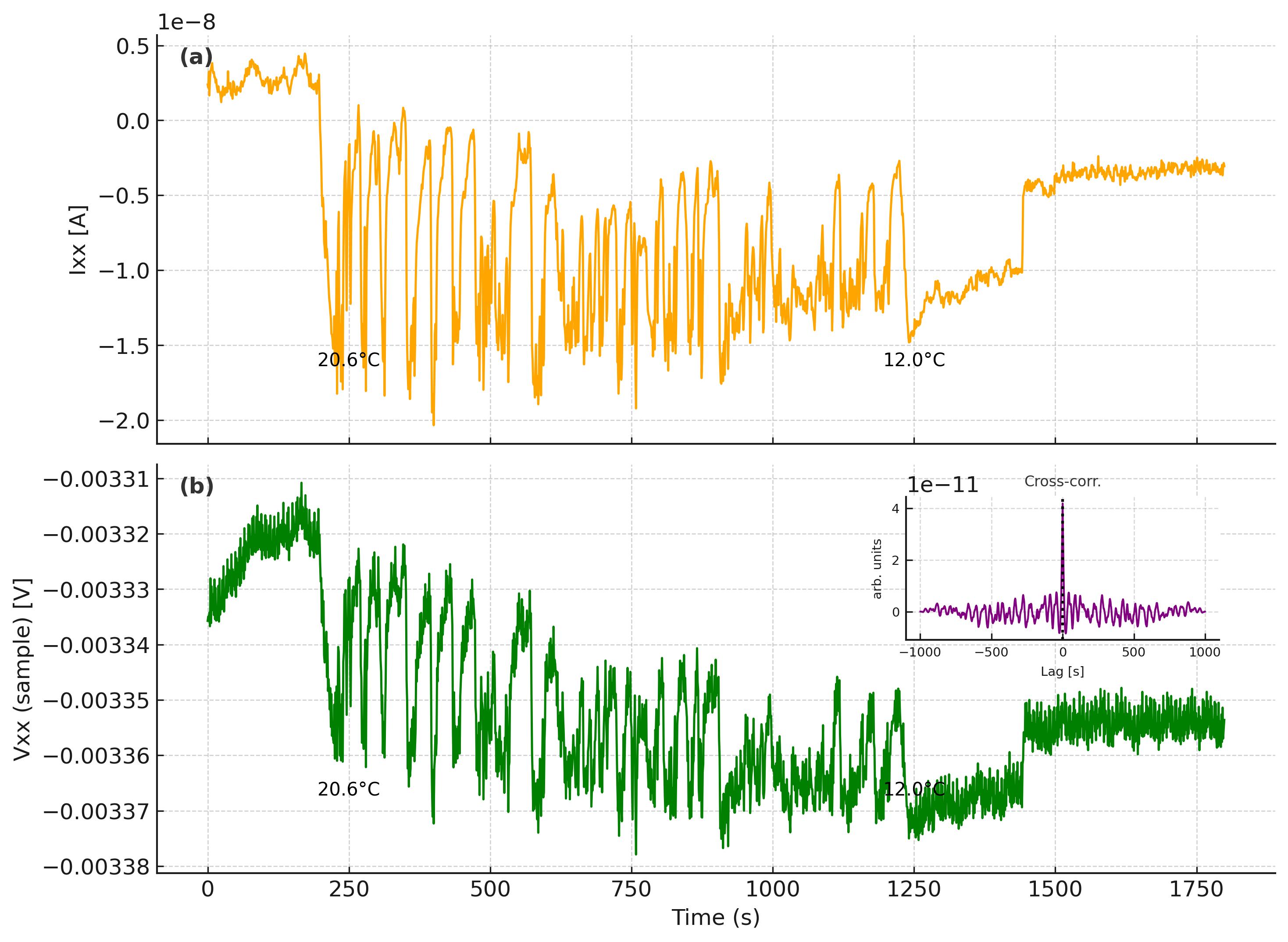}
  \caption{Temperature-driven reorganization of the longitudinal response in hydrated DNA.
  A continuous temperature sweep from above $20.6\,^\circ$C down to below $12.0\,^\circ$C reveals two clear transition points in the longitudinal channels of a hydrated DNA--water matrix (500~ng/$\mu$L, 5~$\mu$L, pH $\approx 8$, $B=500~\mathrm{mT}$).
  (a) The longitudinal current $I_{xx}=V_{xx}/(1~\mathrm{k}\Omega)$ (shunt) shows the onset of large-amplitude, quasi-periodic fluctuations immediately below $20.6\,^\circ$C. As the temperature decreases further, the system undergoes a second reorganization near $12.0\,^\circ$C, after which the fluctuations collapse into a step-like low-temperature branch.
  (b) The longitudinal sample voltage $V_{xx}(\mathrm{sample})$ exhibits a parallel restructuring, with oscillatory features developing below $20.6\,^\circ$C and reorganizing again at approximately $12.0\,^\circ$C.
  The inset shows the cross-correlation between $I_{xx}$ and the longitudinal sample voltage $V_{xx}$, with a pronounced peak near zero lag, indicating correlated fluctuations in the two channels during cooling. Both channels were acquired in the same measurement run at a nominal sampling rate of 1~Hz.}
  \label{fig:fig2}
\end{figure}

To examine the temperature dependence of the electrical response,
we performed cooling measurements at a fixed magnetic field and
recorded longitudinal and transverse channels at a nominal
sampling rate of $1~\mathrm{Hz}$. Figures~\ref{fig:fig2} and
\ref{fig:temperature_hilbert} present experiments performed on
consecutive days using DNA concentrations of $500$ and
$100~\mathrm{ng}/\mu\mathrm{L}$, respectively, at
$B=500~\mathrm{mT}$. In Figure~\ref{fig:fig2}, the longitudinal current
$I_{xx}$ and sample voltage $V_{xx}$ show changes in fluctuation
amplitude and correlation structure near $20.6$ and
$12.0\,^\circ\mathrm{C}$. In
Figure~\ref{fig:temperature_hilbert}\textbf{A}, the transverse
voltage $V_{xy}$ develops large-amplitude, nearly periodic
oscillations below approximately $12\,^\circ\mathrm{C}$,
with peak-to-peak excursions exceeding $150~\mathrm{mV}$.
The proximity of the lower-temperature changes in these two
measurements motivates comparison of their longitudinal and
transverse dynamics.

Analysis of the numerically reconstructed transverse trace
reveals a dominant spectral peak near $0.058~\mathrm{Hz}$
(Figure~\ref{fig:temperature_hilbert}\textbf{B}).
The complex analytic trajectory of the
$0.035$--$0.075~\mathrm{Hz}$ band-pass-filtered component
executes successive circuits around the origin, with a varying
radius (Figure~\ref{fig:temperature_hilbert}\textbf{C}).
Over $940$--$1160~\mathrm{s}$, the reconstructed unwrapped
Hilbert phase accumulates approximately $12.5$ turns.
A linear fit to its progression yields $R^2\approx0.9995$
and a mean winding frequency of $0.0585~\mathrm{Hz}$,
corresponding to a period of about $17~\mathrm{s}$. Comparable winding is obtained
without band-pass filtering after subtraction of the mean,
indicating that it is not introduced by the selection of the
fundamental component. These approximate values characterize
the reconstructed trace. The low-temperature response therefore
exhibits sustained Hilbert-phase winding at fixed magnetic field,
in addition to the pronounced increase in oscillation amplitude.

\subsubsection*{Temperature-dependent balance of ohmic and polarization contributions}

Near 20~\degC, the electrical response contains both ohmic and polarization contributions. Cooling changes their relative importance, providing a physical context for the longitudinal reorganization near 20.6~\degC\ in Figure~\ref{fig:fig2}. This interpretation connects the present measurements with the ohmic--polarization crossover reported previously~\cite{Pietruszka2026BioSystems}.

A further change occurs near 12~\degC: the longitudinal fluctuations in Figure~\ref{fig:fig2} settle into a lower-amplitude branch, while the separate transverse measurement in Figure~\ref{fig:temperature_hilbert} develops large-amplitude, nearly periodic oscillations. These observations suggest a temperature-dependent redistribution of the electrical response, motivating a description that combines conductive and polarization dynamics.

\subsubsection*{Temperature ranges of the measured response}

\begin{figure}[t]
  \centering
  \includegraphics[width=0.8\linewidth]{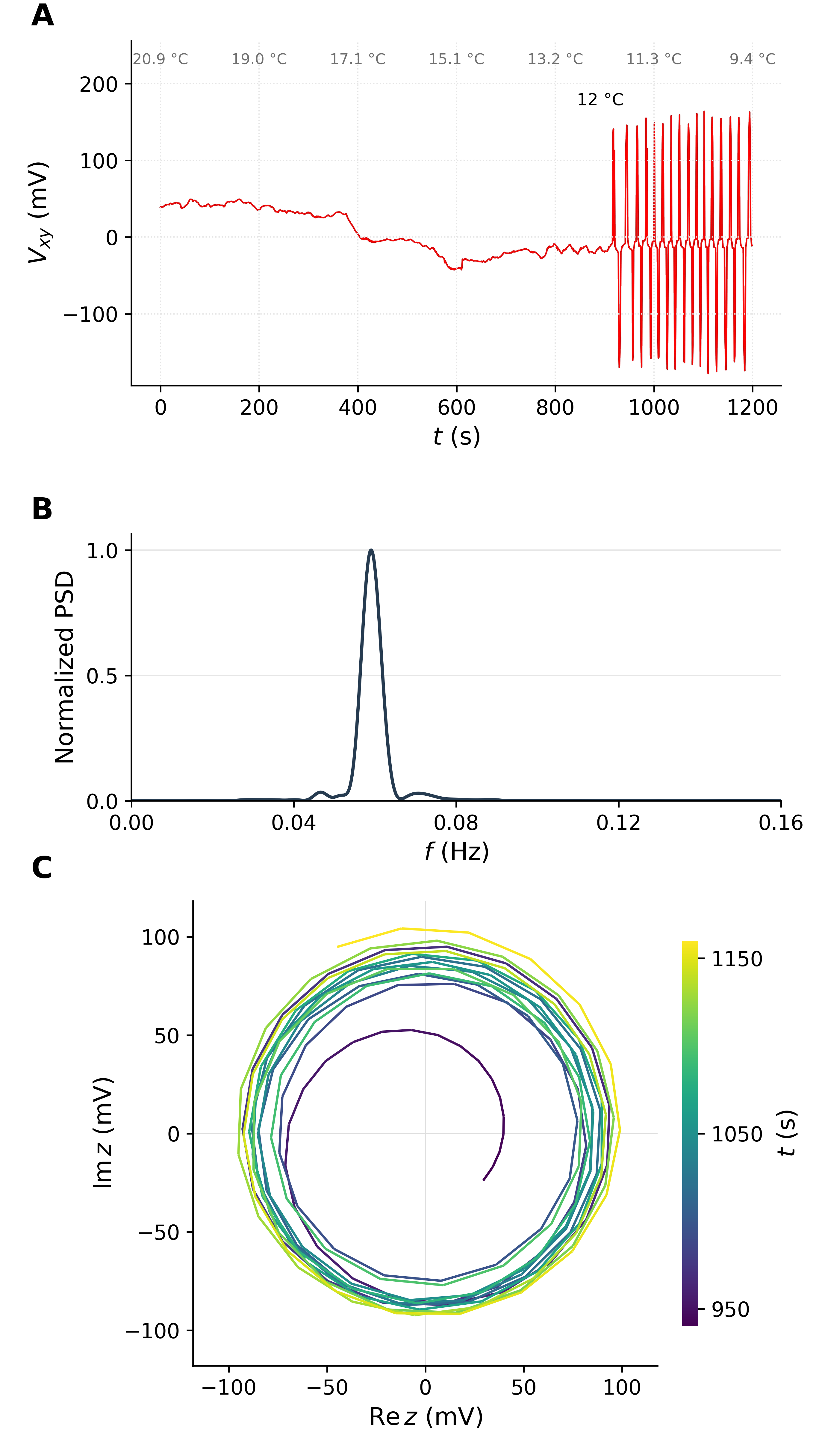}
  \caption{Temperature-driven onset of large-amplitude
  transverse-voltage oscillations and associated Hilbert-phase dynamics.
  \textbf{(A)} A cooling measurement from
  $20.9\,^\circ\mathrm{C}$ to $9.4\,^\circ\mathrm{C}$
  at fixed magnetic field ($B=500~\mathrm{mT}$) records
  $V_{xy}$ in a $100~\mathrm{ng}/\mu\mathrm{L}$ DNA--water sample.
  Above approximately $12\,^\circ\mathrm{C}$, the trace shows
  small fluctuations and slow drift. Below this temperature,
  large-amplitude, nearly periodic oscillations develop,
  with peak-to-peak excursions exceeding $150~\mathrm{mV}$
  and persisting down to $9.4\,^\circ\mathrm{C}$.
  Temperature annotations above the trace mark the progression
  of cooling. Data were recorded at a nominal sampling rate
  of $1~\mathrm{Hz}$.
  \textbf{(B)} Normalized power spectral density of the
  reconstructed oscillatory segment ($920$--$1190~\mathrm{s}$),
  calculated using a Hann window, shows a dominant peak
  near $0.058~\mathrm{Hz}$.
  \textbf{(C)} Complex analytic trajectory
  $z(t)=v_f(t)+i\mathcal{H}[v_f(t)]$ over
  $940$--$1160~\mathrm{s}$, where $v_f(t)$ is the
  $0.035$--$0.075~\mathrm{Hz}$ band-pass-filtered component
  and $\mathcal{H}$ denotes the Hilbert transform.
  Color indicates time. Successive circuits around the
  origin reveal sustained phase winding with a varying radius.}
  \label{fig:temperature_hilbert}
\end{figure}

The cooling measurements distinguish three descriptive temperature ranges. Above approximately 20.6~\degC, the longitudinal response precedes the first pronounced reorganization. Between approximately 20.6 and 12~\degC, the longitudinal channels exhibit enhanced, structured fluctuations. Below approximately 12~\degC, the longitudinal measurement enters a lower-amplitude branch, while the separate transverse experiment displays sustained, large-amplitude oscillations. These ranges summarize the measurements in Figures~\ref{fig:fig2} and \ref{fig:temperature_hilbert} and provide a basis for examining how cooling modifies the balance between ohmic and polarization contributions.

\subsubsection*{Magnetic-field dependence at fixed temperature}

\begin{figure}[t]
  \centering
  \includegraphics[width=\linewidth]{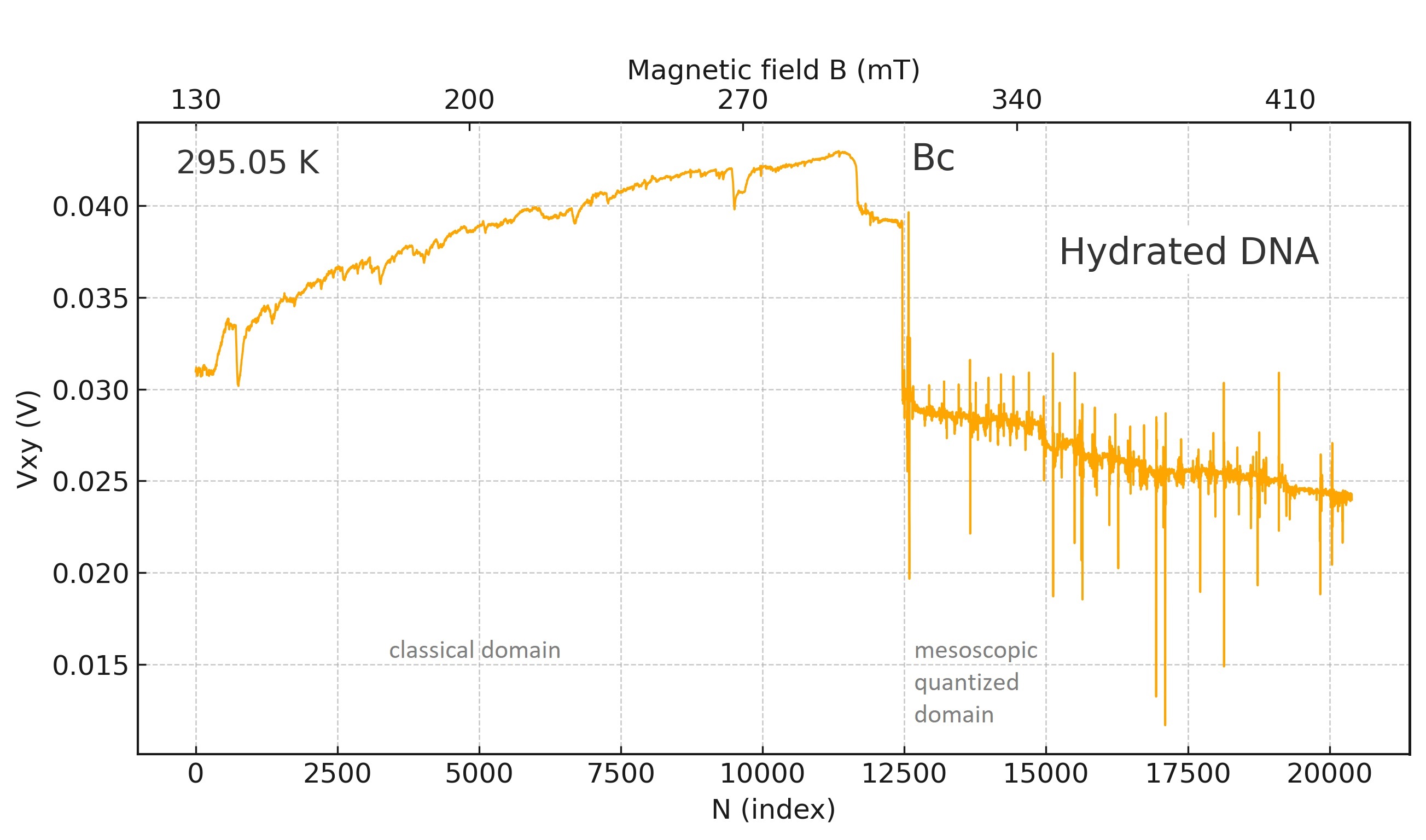}
  \caption{Initial observation of a sharp transverse-voltage transition followed by a staircase-like response in hydrated DNA. The trace was recorded for a 500~ng/$\mu$L genomic DNA sample (pH $\approx8$, $T=295.05$~K, $21.9\,^\circ\mathrm{C}$). The lower axis shows the acquisition index $N$, and the upper axis gives the approximate magnetic-field calibration, from 130~mT at $N=0$ to 420~mT at $N\approx20\,000$. Near $B_c$, $V_{xy}$ drops from approximately 0.039 to 0.029~V, followed by a descending sequence of plateau-like levels. In this first experiment, the magnet position was adjusted manually during acquisition, producing localized transients alongside background fluctuations of approximately $\pm0.1$~mV. The trace preserves the original observation that motivated subsequent investigations of thresholds and discrete voltage structures. The raw dataset is available on Zenodo. The underlying experimental data were previously published in Physica B~\cite{Pietruszka2025a}; the present figure provides a new graphical representation of those data within the polarization-based interpretation developed here.}
  \label{fig:fig4}
\end{figure}

We next examine the transverse-voltage response during a magnetic-field sweep at near-room temperature. For a DNA concentration of 500~ng/$\mu$L (Figure~\ref{fig:fig4}), \Vxy\ exhibits a sharp threshold at $B_c$, followed by a staircase-like sequence of plateau levels. The approximately regular spacing of these levels is visible against background fluctuations estimated at $\pm0.1$~mV. The plateau centers and their spacing provide quantitative descriptors of the response, which can be compared with the characteristic scales extracted from the oscillatory measurements discussed below. Within the proposed dynamical framework, the plateaus are interpreted as intervals of residence in metastable configurations.

Pronounced fluctuations occur near several boundaries between neighboring plateaus, followed by intervals of relatively stable voltage. In this initial experiment, which revealed the sharp transition and subsequent voltage staircase, the magnet position was adjusted manually; localized transients accompany these adjustments in Figure~\ref{fig:fig4}. Their temporal relationship to the voltage steps is therefore relevant to interpreting the transition dynamics.

Within a metastable-state description, enhanced fluctuations near a switching boundary may accompany changes in the stability of neighboring configurations. Fluctuation-assisted transitions are familiar in nonlinear systems~\cite{CrossHohenberg1993} and provide a possible framework for examining the observed sequence of plateaus and switching events. Quantitative analysis should distinguish plateau fluctuations from transients associated with adjustments of the applied field.


\subsubsection*{Inverse-field oscillatory structures}

\begin{figure}[t]
  \centering
  \includegraphics[width=\linewidth]{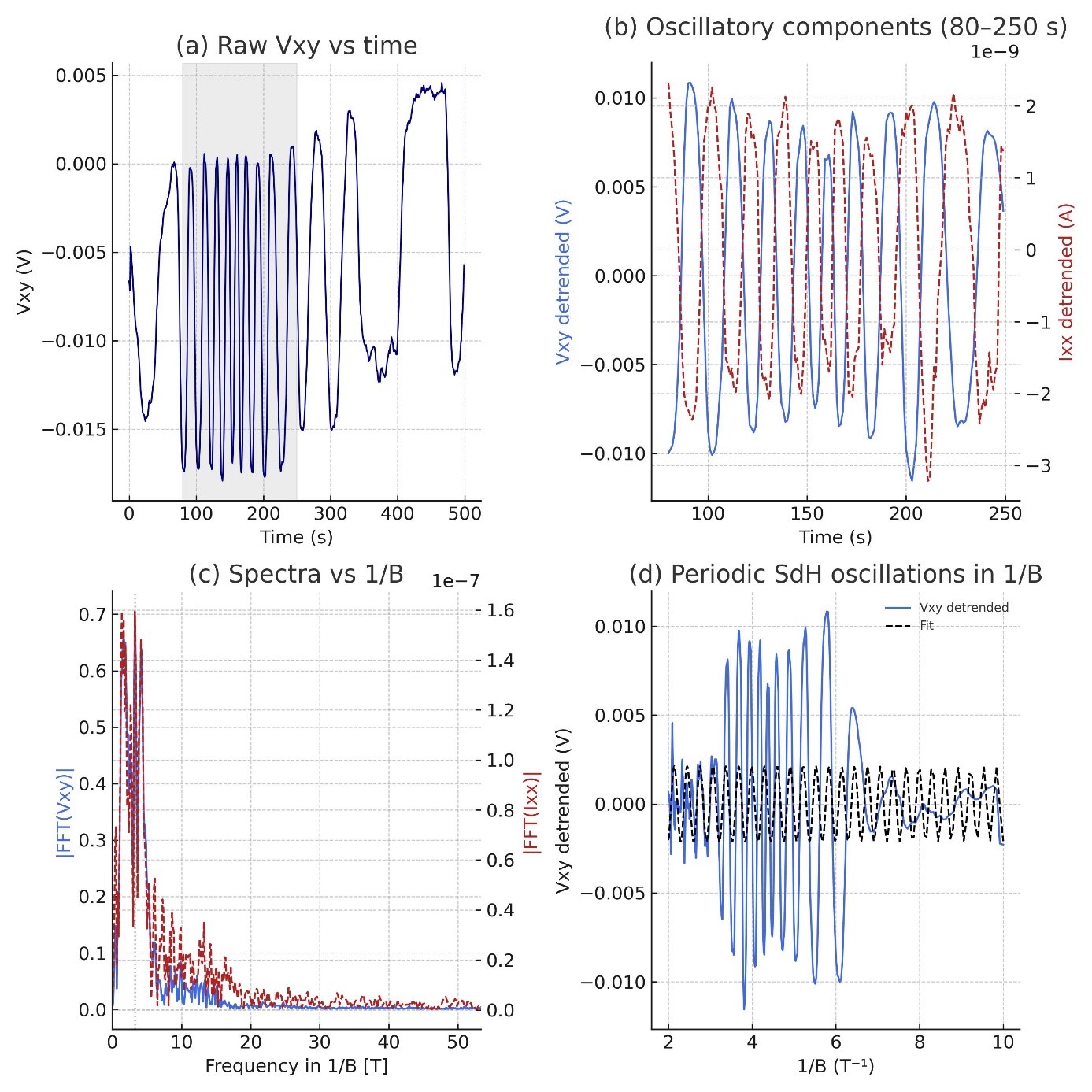}
  \caption{Inverse-field oscillatory structures in longitudinal and transverse channels (100~ng/$\mu$L, $5\,^\circ\mathrm{C}$).
(a) Transverse voltage $V_{xy}$ recorded in a Greek-cross geometry during a continuous magnetic-field sweep from 0.10 to 0.50~T over 500~s.
(b) Detrended transverse voltage and longitudinal current over the 80--250~s interval, displayed on separate axes.
(c) FFT analysis in the $1/B$ domain yields closely matching frequencies, $F_{V_{xy}}\approx6.31$~T and $F_{I_{xx}}\approx6.27$~T, with a reported spacing $\Delta(1/B)=0.159\pm0.002$~T$^{-1}$.
(d) Detrended transverse voltage plotted versus $1/B$, together with the fitted oscillatory component. The transport-inspired parameterization gives $A_F=(6.0\pm0.1)\times10^{16}$~m$^{-2}$ and $k_F=(1.38\pm0.01)\times10^8$~m$^{-1}$, treated here as effective derived scales.}
  \label{fig:fig5}
\end{figure}

For a DNA concentration of 100~ng/$\mu$L at 5~\degC\ (Figure~\ref{fig:fig5}), the transverse and longitudinal channels exhibit oscillatory structures over the magnetic-field range 0.13--0.30~T. Analysis in the inverse-field domain yields closely matching characteristic frequencies, $F_{V_{xy}}\approx6.31$~T and $F_{I_{xx}}\approx6.27$~T, corresponding to a reported spacing $\Delta(1/B)=0.159\pm0.002$~T$^{-1}$. Their agreement suggests a shared field-dependent modulation of the two responses. The parameters \AF\ and \kF\ shown in Figure~\ref{fig:fig5} are derived through the transport-inspired parameterization specified in Methods and serve as effective scales for comparing the observed structures.

\begin{figure}[t]
  \centering
  \includegraphics[width=\linewidth]{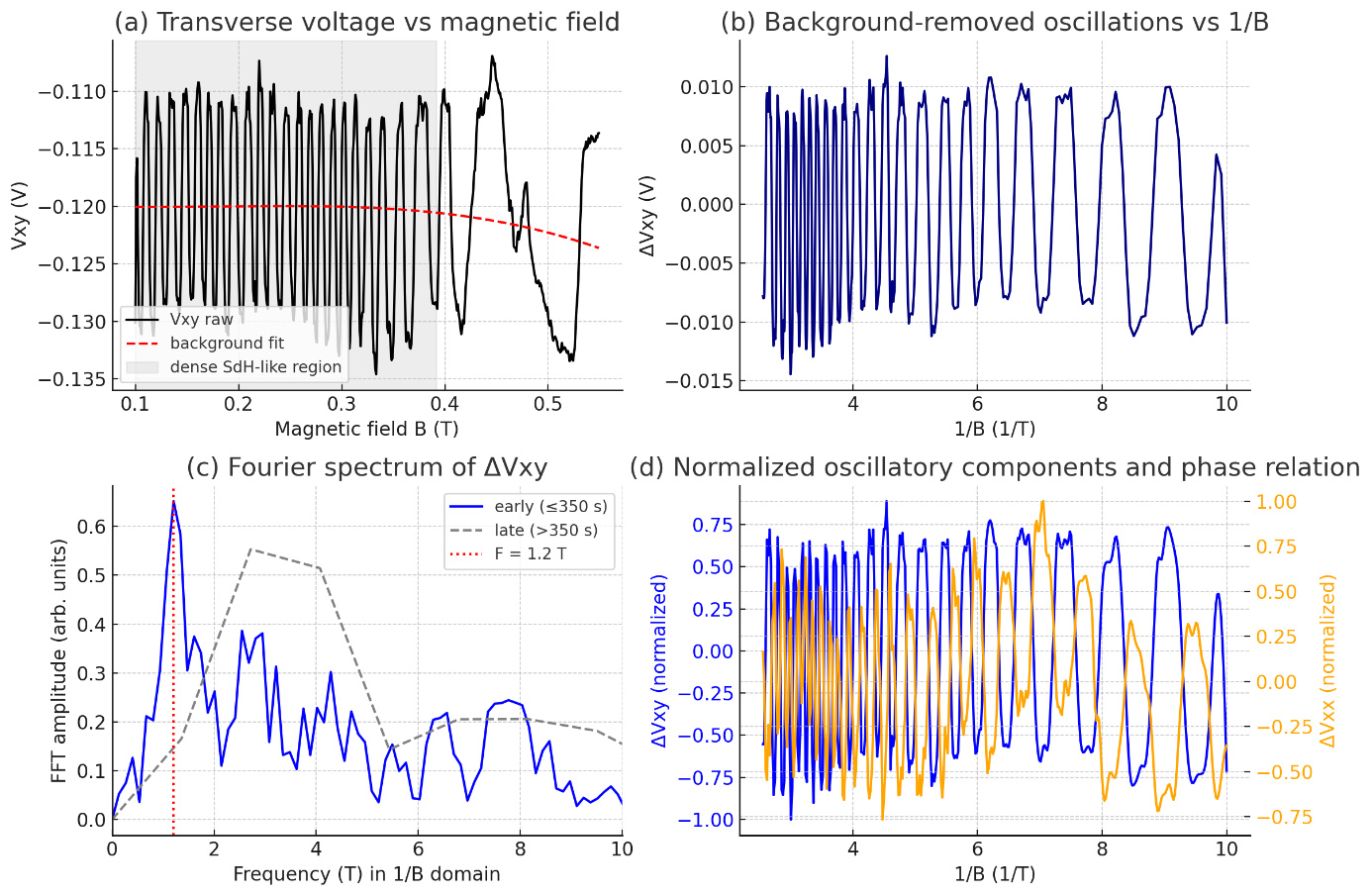}
  \caption{Inverse-field oscillatory structures at $6\,^\circ\mathrm{C}$ (500~ng/$\mu$L).
(a) Transverse voltage $V_{xy}$ during a 540~s magnetic-field sweep from 0.10 to 0.55~T (relative humidity 29\%). The raw signal (black) and fitted polynomial background (red dashed) are shown.
(b) Background-subtracted $\Delta V_{xy}$ plotted versus $1/B$, with approximately periodic modulation and a characteristic spacing near 0.83~T$^{-1}$.
(c) FFT analysis of the early ($t\le350$~s, blue) and late ($t>350$~s, grey) intervals. The dominant inverse-field frequency is approximately $F=1.2$~T; the adopted Onsager-type parameterization gives an effective density scale $n_{\mathrm{eff}}\approx5.8\times10^{14}$~m$^{-2}$.
(d) Normalized transverse and longitudinal modulations, $\Delta V_{xy}$ (blue) and $\Delta V_{xx}$ (orange), displayed on separate axes. Their amplitudes before normalization differ by approximately four orders of magnitude, with a reported relative phase near $\pi$. Above approximately 0.40~T, the transverse response evolves into broader lobes.}
  \label{fig:fig6}
\end{figure}

At 6~\degC\ and 500~ng/$\mu$L (Figure~\ref{fig:fig6}), the background-subtracted transverse signal \dVxy\ exhibits approximately periodic modulation in the inverse-field domain, with a characteristic spacing near 0.83~T$^{-1}$ and a dominant FFT peak at $F\approx1.2$~T. The adopted Onsager-type parameterization yields an effective density scale $\neff\approx5.8\times10^{14}\,\mathrm{m}^{-2}$, comparable to the plateau-derived estimate. In the analyzed interval, the transverse modulation exceeds the longitudinal modulation by approximately four orders of magnitude, with a reported relative phase near $\pi$. For $B\gtrsim0.4$~T, the oscillatory pattern evolves into broader lobes, marking a change in the measured response.

Complementary measurements were performed with the mechanical drive switched off and the magnetic field held fixed while the sample cooled passively. Under these conditions, large-amplitude temporal oscillations persisted, as illustrated by the cooling measurement in Figure~\ref{fig:temperature_hilbert}. This observation establishes that oscillatory responses also occur under stationary magnetic-field conditions and provides a complementary comparison with the field-sweep experiments.

\subsubsection*{Tensor-based representation of the measured response}

\begin{figure}[t]
  \centering
  \includegraphics[width=\linewidth]{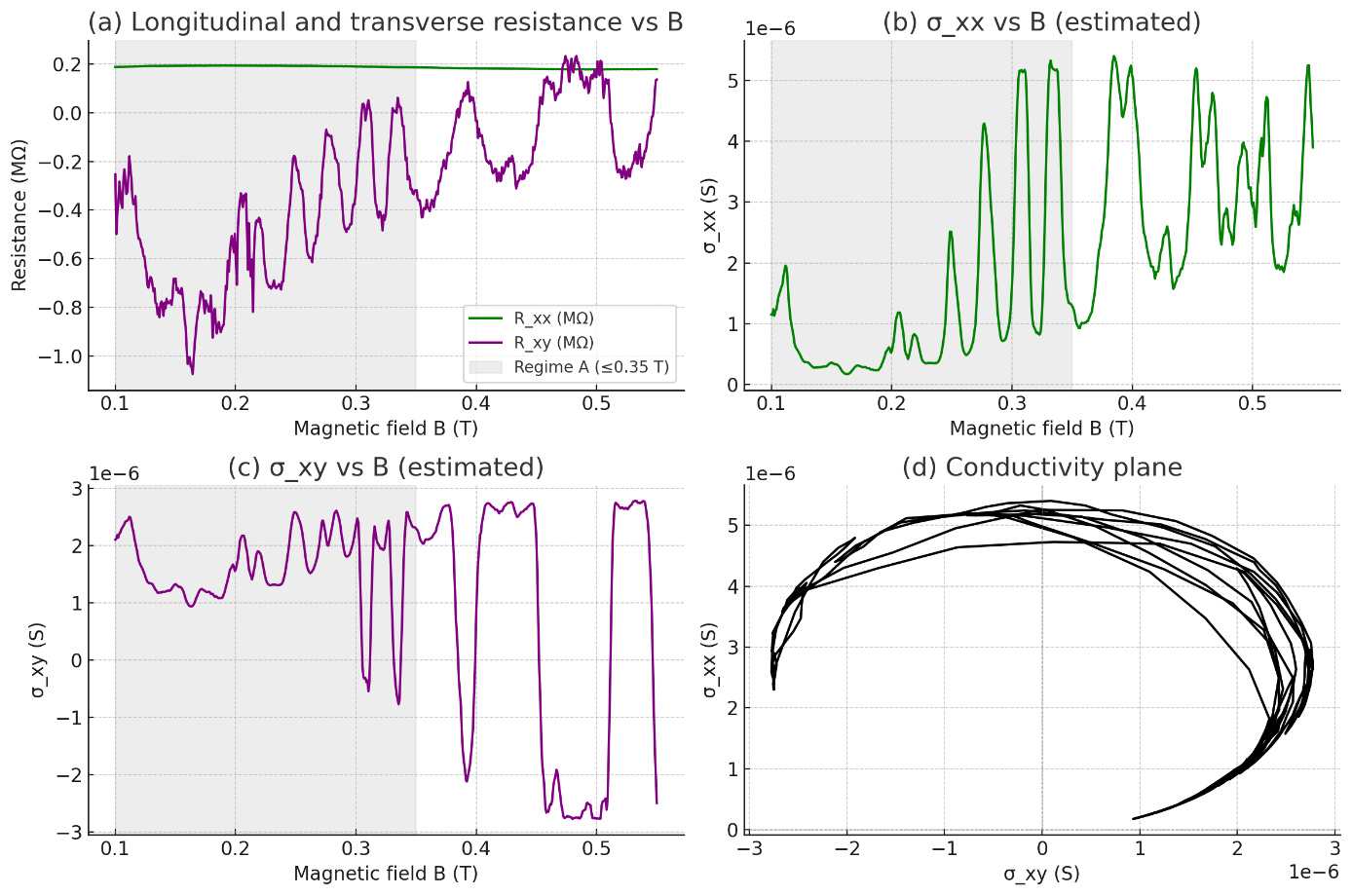}
  \caption{Field-dependent response ratios and tensor-based representation.
(a) Effective ratios $R_{xx}$ and $R_{xy}$ calculated from the measured voltages using the current profile. For $B\lesssim0.35$~T, the curves display oscillatory structure in the inverse-field domain.
(b,c) Derived quantities $\sigma_{xx}(B)$ and $\sigma_{xy}(B)$ obtained by tensor inversion within the adopted parameterization. Above the threshold near $B_c\approx0.35$~T, the low-field modulation evolves into a broader envelope.
(d) Parametric $\sigma_{xx}$--$\sigma_{xy}$ trajectory illustrating the relationship between the two derived response components during the field sweep.}
  \label{fig:fig7}
\end{figure}

Figure~\ref{fig:fig7} presents effective response ratios $R_{xx}(B)$ and $R_{xy}(B)$ obtained from the measured voltages and current profile, together with the quantities $\sigma_{xx}(B)$ and $\sigma_{xy}(B)$ calculated by tensor inversion. This representation provides a compact comparison of the longitudinal and transverse responses within the adopted parameterization. The resulting field-dependent curves and parametric trajectories display structured variations, with the low-field modulation evolving into a broader envelope above the threshold. We use these derived quantities to characterize the change in response and relate it to the phenomenological dynamical framework.

\subsubsection*{Control comparisons and sample preparation}

\begin{figure}[t]
  \centering
  \includegraphics[width=\linewidth]{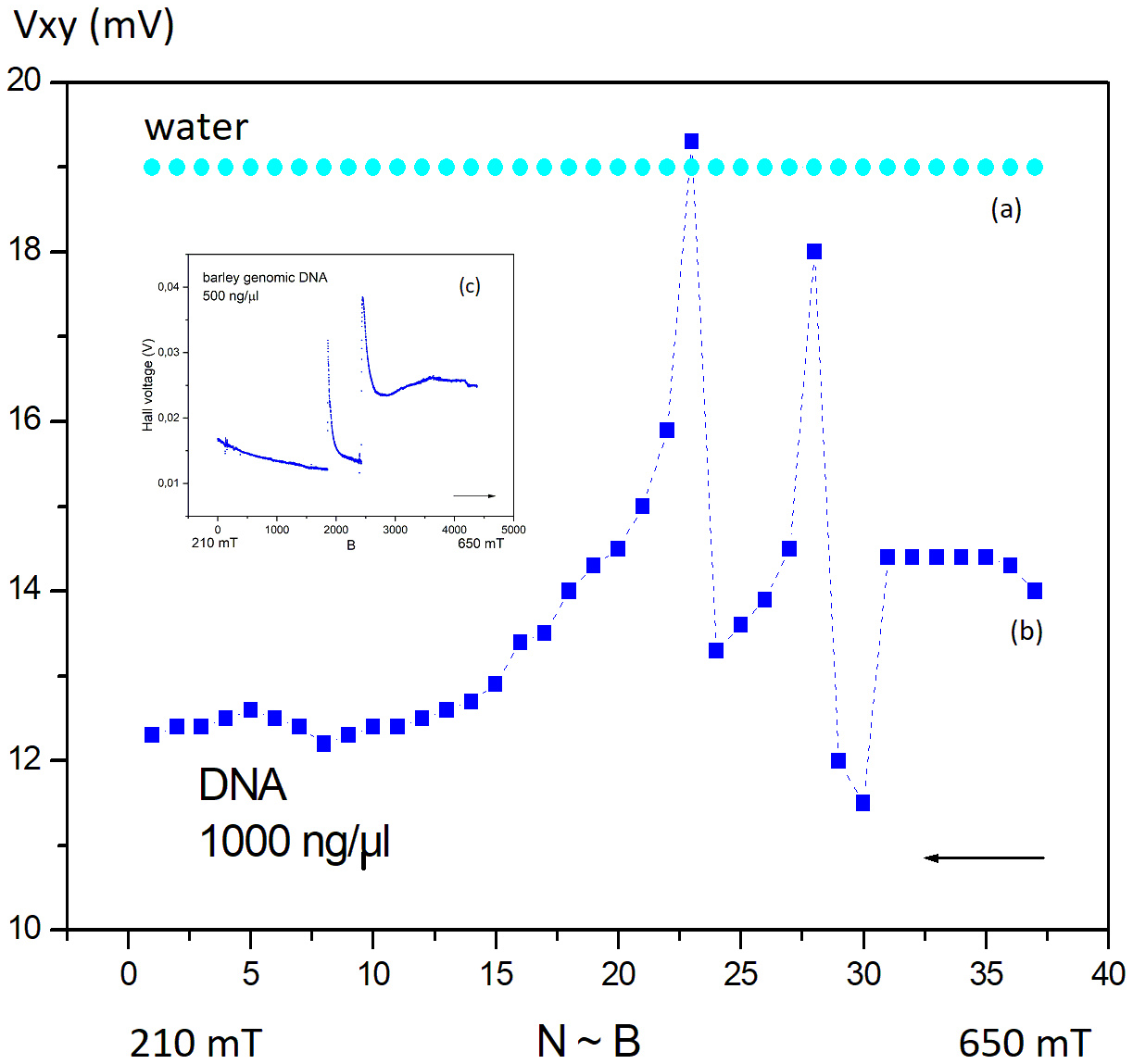}
  \caption{Comparison of distilled-water and hydrated-DNA responses.
(a) Transverse voltage $V_{xy}$ recorded for distilled water, providing a reference without the pronounced structures seen in the DNA traces.
(b,c) DNA--water samples at 1000 and 500~ng/$\mu$L exhibit a double-peak motif followed by plateau-like levels. Arrows indicate the magnetic-field sweep directions. The principal features occur at similar field positions in the forward and reverse sweeps, showing approximate agreement within the resolution of the displayed measurements.}
  \label{fig:fig8}
\end{figure}

The distilled-water trace in Figure~\ref{fig:fig8}a provides a reference without the pronounced thresholds, oscillatory patterns, and plateau-like structures observed in the corresponding DNA measurements. Samples at 1000 and 500~ng/$\mu$L (Figures~\ref{fig:fig8}b,c) exhibit a double-peak motif followed by plateau-like levels. The principal features occur at similar field positions in both sweep directions, indicating approximate sweep-direction agreement within the resolution of the displayed measurements. These comparisons support an association between the structured response and the DNA-containing samples. 

\begin{figure}[t]
  \centering
  \includegraphics[width=\linewidth]{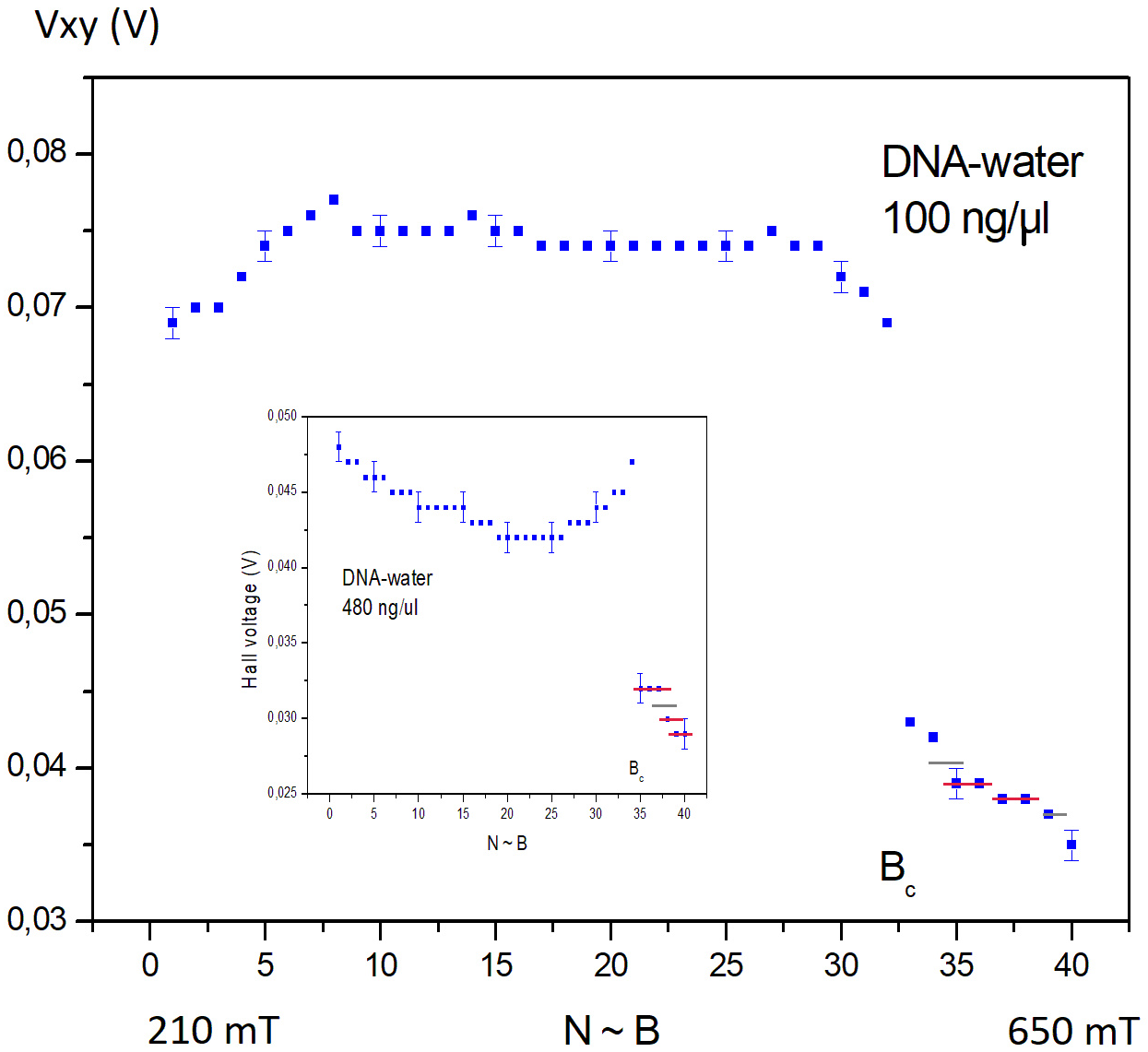}
  \caption{Threshold and plateau-like voltage structures in DNA prepared without added buffer.
Main panel: transverse voltage $V_{xy}$ for a 100~ng/$\mu$L DNA sample dissolved directly in distilled water.
Inset: a 480~ng/$\mu$L sample prepared under the same buffer-free conditions.
Both traces exhibit a threshold near $B\approx0.6$~T, with a voltage drop exceeding 25~mV, followed by weakly resolved plateau-like levels with characteristic spacing near 1~mV. Red lines mark the most clearly identified levels; grey lines indicate less clearly resolved levels. The comparison documents similar threshold and voltage-spacing scales across the two concentrations under the displayed conditions.}
  \label{fig:fig9}
\end{figure}

To examine the influence of preparation conditions, we measured DNA dissolved directly in distilled water (Figure~\ref{fig:fig9}). Both the 100 and 480~ng/$\mu$L samples exhibit a pronounced threshold near $B\approx0.6$~T, with a voltage drop exceeding 25~mV, followed by weakly resolved plateau-like levels. Their characteristic spacing is approximately 1~mV, comparable to the previously reported estimate of 0.8~mV. These observations show that the threshold and staircase-like response also occur in samples prepared without added TE buffer. Metastable transverse-voltage switching has also been reported for barley genomic DNA dissolved in deionized water~\cite{Pietruszka2026SciRep}. A Cu-coated control gives a transverse response of opposite sign, providing an additional comparison with the DNA measurements.

\subsubsection*{Phase-resolved reconstruction of longitudinal and transverse signals}

\begin{figure}[t]
  \centering
  \includegraphics[width=\linewidth]{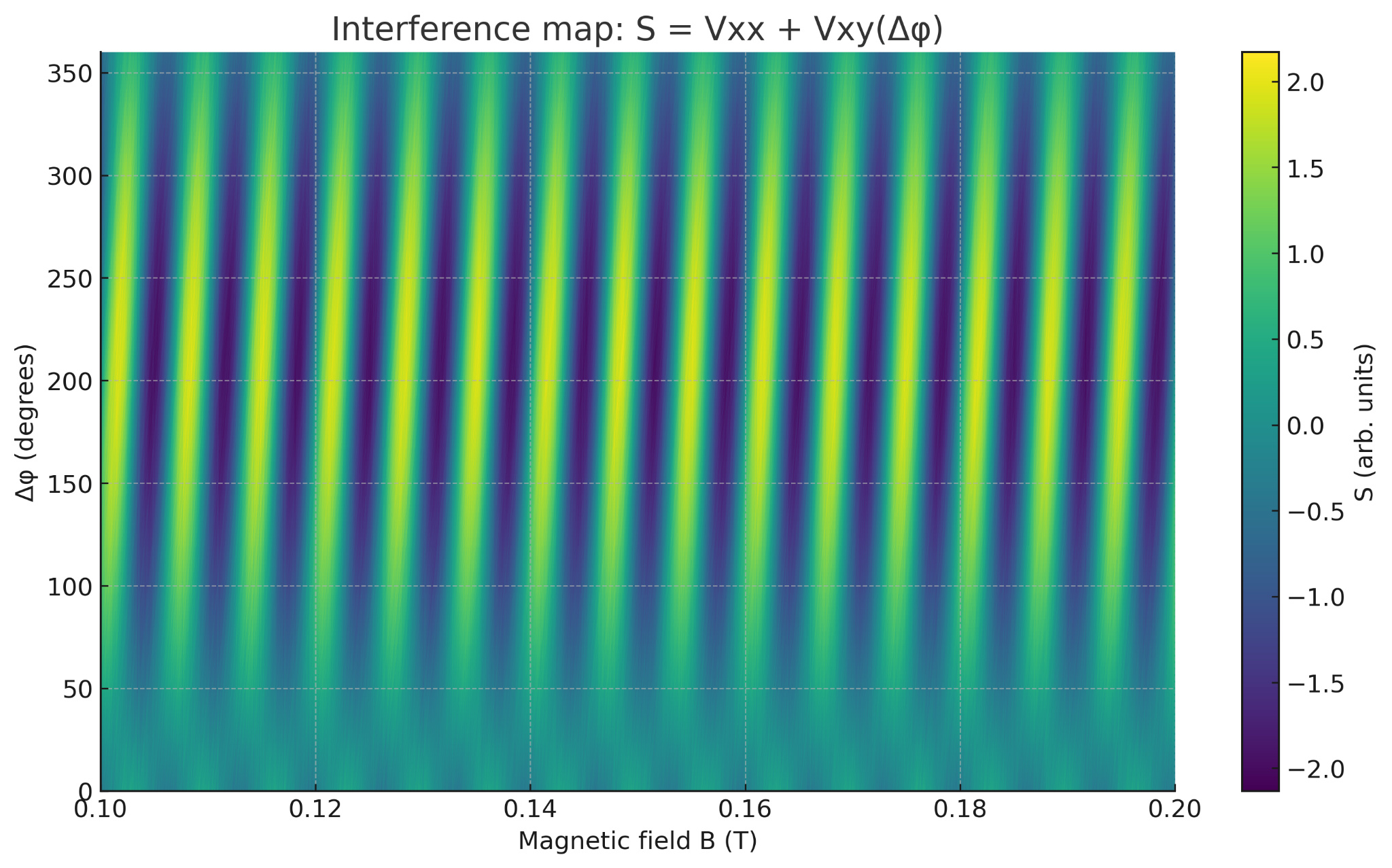}
  \caption{Phase-dependent numerical superposition of longitudinal and transverse signals.
The map $S(t,\Delta\phi)=V_{xx}(t)+V_{xy}(t+\tau)$ is constructed from channels recorded in the same measurement run during a magnetic-field ramp at 10~\degC\ for a 100~ng/$\mu$L DNA sample. For a narrowband component, the imposed time shift $\tau$ corresponds approximately to $\Delta\phi=\omega\tau$. Alternating regions of enhanced and reduced amplitude illustrate the dependence of the summed signal on the imposed shift, with reported maximum contrast near $\Delta\phi\approx200^\circ$. The acquisition introduces a sequential sampling offset of approximately 0.25~s between channels, which must be accounted for when comparing their relative phases.}
  \label{fig:fig10}
\end{figure}

Figure~\ref{fig:fig10} presents a numerical reconstruction obtained by summing the longitudinal signal \Vxx$(t)$ and the time-shifted transverse signal \Vxy$(t+\tau)$ recorded in the same measurement run. For a 100~ng/$\mu$L DNA sample at 10~\degC\ during a magnetic-field ramp over 100--200~mT, the map displays alternating regions of enhanced and reduced amplitude. These patterns characterize how the recorded oscillatory components combine as their relative time shift is varied.

For a narrowband component of angular frequency $\omega$, the imposed shift corresponds approximately to a phase offset $\Delta\phi=\omega\tau$. The reported maximum contrast near $\Delta\phi\approx200^\circ$ provides a descriptor for comparison with the relative phase extracted from the individual channels. The reconstruction offers a visual representation of their phase-dependent relationship, motivating quantitative analysis of phase stability and correlation within the proposed dynamical framework.


\subsubsection*{Relationship between transverse voltage and photovoltage}

\begin{figure}[t]
  \centering
  \includegraphics[width=\linewidth]{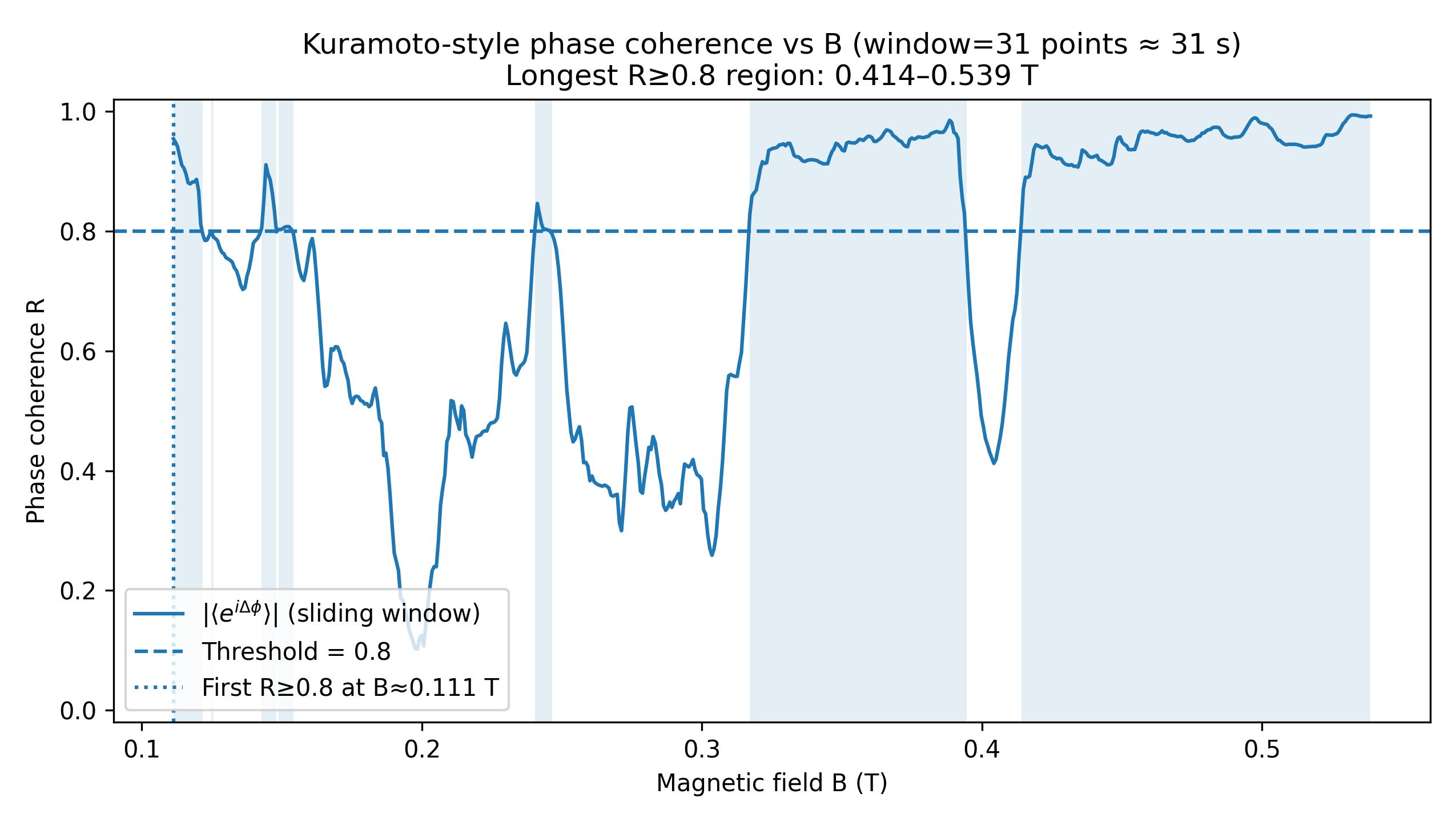}
 \caption{Sliding-window phase agreement between transverse voltage and photovoltage. The plotted quantity is $R=\left|\langle e^{i\Delta\phi}\rangle_w\right|$,
where $\Delta\phi$ is the relative analytic phase and the average is taken over a window of 31 samples (approximately 31~s). The dashed line marks $R=0.8$, and shaded intervals identify values above this threshold. The longest displayed interval with $R\ge0.8$ extends approximately from 0.414 to 0.539~T.}
  \label{fig:fig11}
\end{figure}

Figure~\ref{fig:fig11} shows sliding-window phase agreement between the transverse voltage and photovoltage. Extended intervals with $R\ge0.8$ characterize persistent relative-phase organization within the analyzed signals. This representation provides a complementary view of the relationship between the two recorded signals. The photovoltaic sensor was spatially separated from the sample contacts and had no galvanic connection to the sample circuit. Possible coupling pathways include electromagnetic, thermal, and other influences within the shared experimental environment.

\begin{figure}[t]
  \centering
  \includegraphics[width=\linewidth]{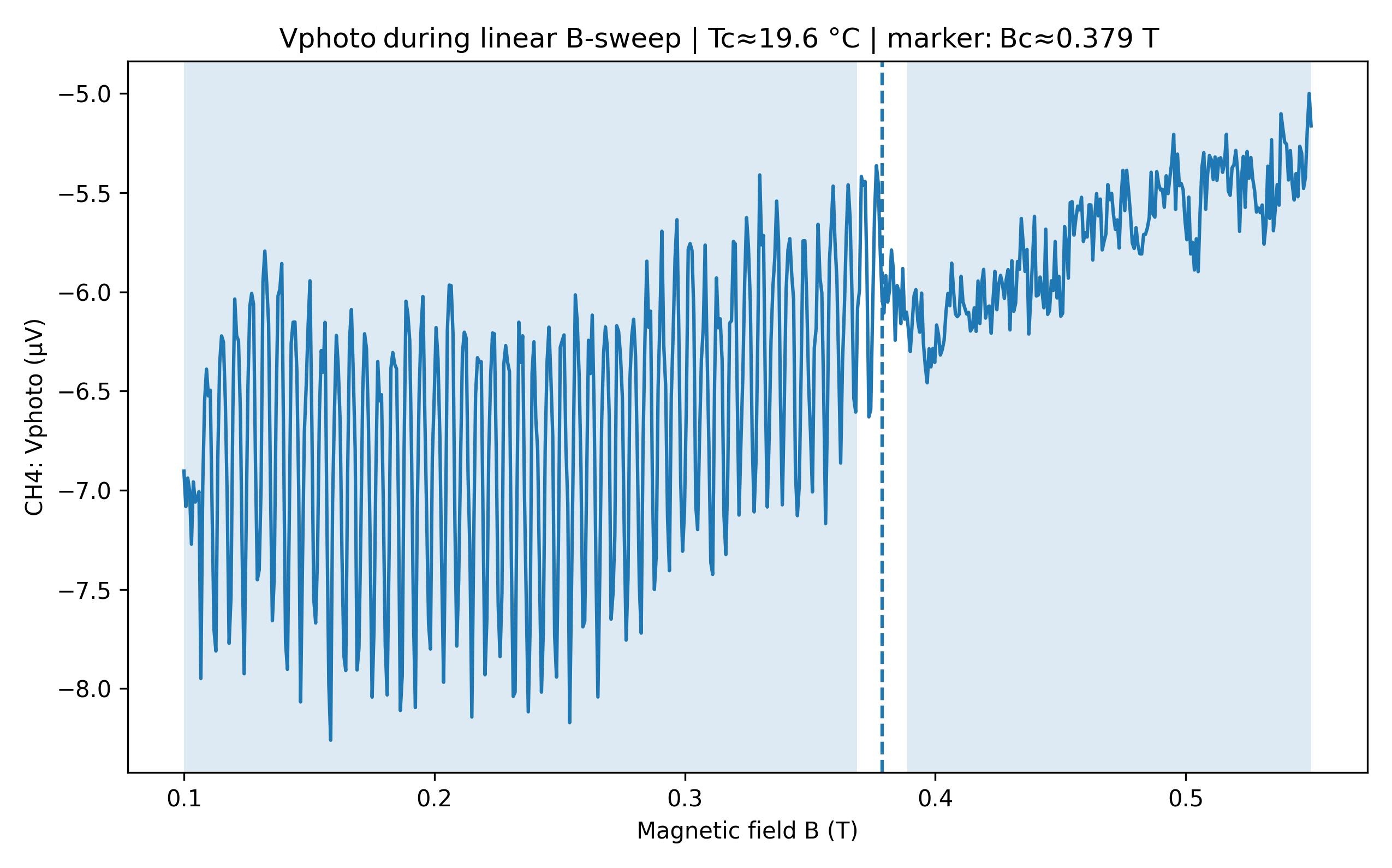}
  \caption{Photovoltaic response during a magnetic-field sweep.
The auxiliary photovoltage (CH4, $V_{\mathrm{PV}}$) was recorded from a photovoltaic plate positioned approximately 3~cm from a hydrated DNA sample (5~\uL, 500~ng/$\mu$L). The magnetic field was swept from approximately 0.10 to 0.55~T over 600~s, with a nominal sampling interval of 1~s. The temperature was approximately 19.6~\degC\ and the relative humidity approximately 23\%, with a 0.1~V bias applied to the sample circuit. Electrical channels CH1--CH3 were recorded in the same run. Near $t\approx371$~s, the photovoltage changes from a densely oscillating, higher-amplitude response to a lower-amplitude regime. Assuming a linear field sweep, this transition corresponds to $B\approx0.38$~T.}
  \label{fig:fig12}
\end{figure}

Figure~\ref{fig:fig12} shows the photovoltage recorded during a magnetic-field sweep alongside the electrical channels. Near $t\approx371$~s, the photovoltage changes from a densely oscillating, higher-amplitude response to a lower-amplitude regime. Under the assumed linear field calibration, this change corresponds to $B\approx0.38$~T. The transition in the spatially separated photovoltaic channel provides an additional observable for comparison with the electrical response and for investigating the coupling between the sample, detector, and experimental environment.


\subsubsection*{Comparison of effective density scales}

Within the adopted parameterization, the inverse-field oscillation analysis gives
\[
\neff\approx(4.8\pm0.5)\times10^{14}\,\mathrm{m}^{-2},
\]
while the plateau analysis yields
\[
\neff\approx(4.69\pm0.5)\times10^{14}\,\mathrm{m}^{-2}.
\]
The two effective density estimates arise from the oscillatory and plateau analyses presented here. Their quantitative comparison depends on the conversion relations and uncertainty assumptions used in each analysis.

\subsubsection*{Schematic field--temperature response map}

\begin{figure}[t]
  \centering
  \includegraphics[width=\linewidth]{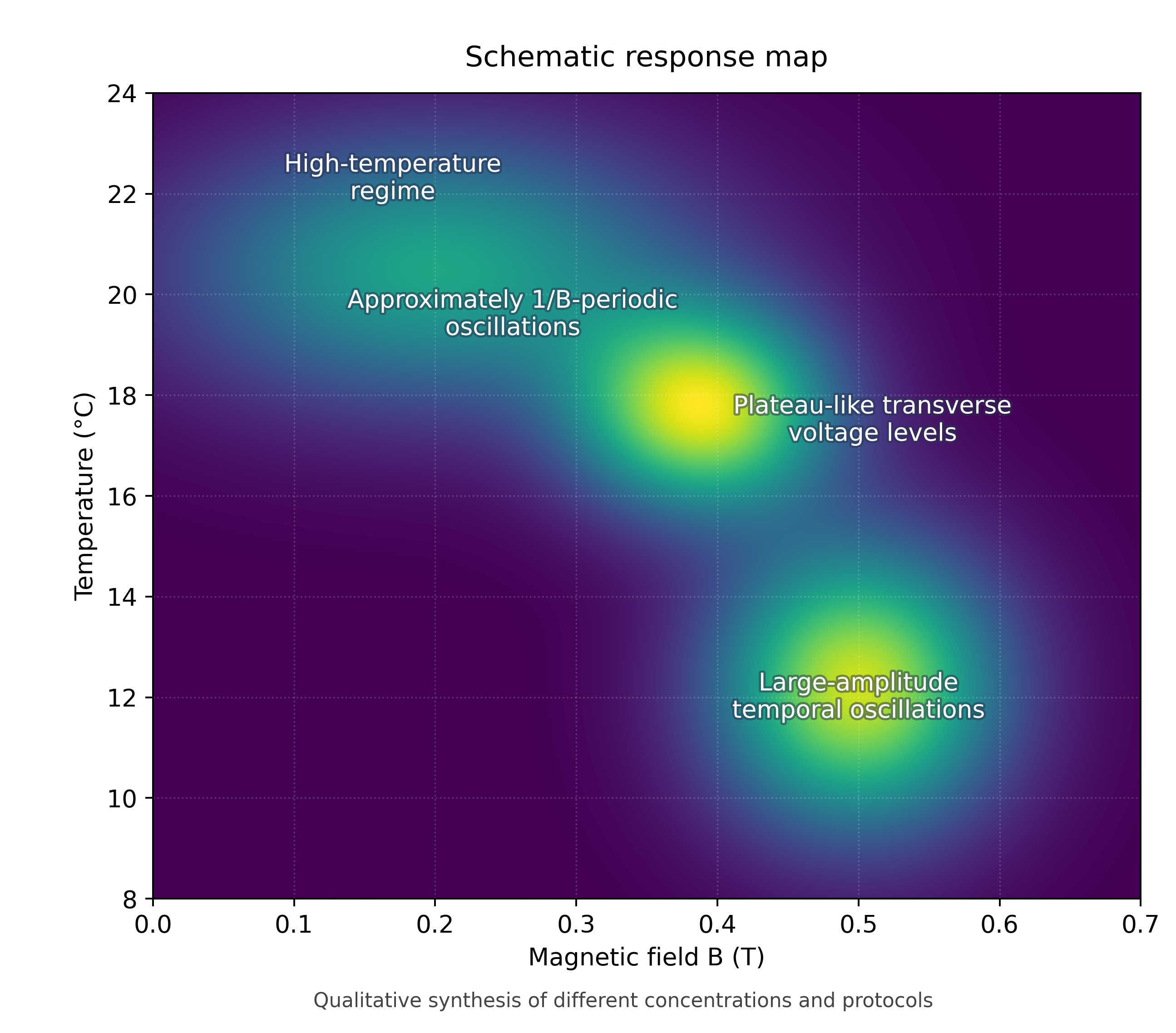}
  \caption{Schematic magnetic-field--temperature response map for hydrated DNA. Labels indicate a high-temperature regime, approximately $1/B$-periodic oscillations, plateau-like transverse-voltage levels, and large-amplitude temporal oscillations during cooling. The map provides a qualitative synthesis of observations obtained at different DNA concentrations and under different measurement protocols. The positions and extents of the colored regions are illustrative; colors do not represent a measured quantity, and the map does not establish phase boundaries or assign microscopic mechanisms. The schematic was prepared with AI assistance.}
  \label{fig:fig13}
\end{figure}

Figure~\ref{fig:fig13} provides a schematic overview of selected response regimes discussed above: a high-temperature regime, approximately inverse-field-periodic oscillations, plateau-like transverse-voltage levels, and large-amplitude temporal oscillations during cooling. It combines observations obtained at different DNA concentrations and under different measurement protocols. The colored regions are illustrative and do not represent a measured quantity or experimentally determined phase boundaries.


The longitudinal measurement provides complementary information about the balance between conductive and polarization contributions. During cooling at fixed magnetic field, the current decreases sharply and subsequently fluctuates near zero in the displayed interval. Together with the mixed ohmic and polarization response near 20~\degC, this behavior suggests a temperature-dependent change in the relative contributions to the measured signal. A related longitudinal crossover has been reported previously~\cite{Pietruszka2026BioSystems}. The present longitudinal and transverse observations motivate a dynamical description that accommodates both contributions and their evolution with temperature.

\section{Discussion}
The transverse voltage in hydrated DNA exhibits field-dependent thresholds, staircase-like plateaus, and oscillations approximately periodic in $1/B$. During cooling at fixed magnetic field, the transverse signal develops sustained phase winding below approximately $12\,^\circ\mathrm{C}$. In a separate field sweep, $V_{xy}$ and the signal from a photovoltaic plate positioned approximately 3~cm from the sample show extended intervals of relative-phase agreement. Together with the previously reported suppression of longitudinal current during cooling~\cite{Pietruszka2026BioSystems} and metastable transverse-voltage switching~\cite{Pietruszka2026SciRep}, these observations motivate a phenomenological description in terms of collective polarization dynamics.

\subsubsection*{Polarization dynamics and metastable response levels}

The coexistence of conductive and polarization contributions provides a framework for interpreting the electrical response across temperature regimes. Within this framework, the staircase-like voltage sequence may reflect residence in metastable configurations, with changes in magnetic field modifying their stability. Multistability and fluctuation-assisted switching are established concepts in driven nonlinear systems~\cite{CrossHohenberg1993} and motivate the minimal dynamical model developed below.

A quantitative description requires a relation between the collective state variable and the measured voltage. Polarization dynamics can contribute through time-dependent charge redistribution and the resulting electrode potentials, while conductive pathways and circuit impedance influence the detected signal. This connection is central to explaining both the plateau levels and the transitions between them.

Under a constant applied voltage, charging of the electrical double layers at ideally blocking electrodes screens the bulk electric field as the electrolyte approaches equilibrium, and the charging current decays~\cite{Bazant2004}. In our setup, the 0.1~V is the bias applied to the sample circuit. The voltage drop across the liquid bulk was not determined separately from the drops at the copper--electrolyte interfaces, and the applied bias alone does not establish whether the contacts were blocking. During magnetic-field or temperature sweeps, time-dependent charge redistribution in the liquid or at the interfaces can contribute to the measured transverse voltage even when the longitudinal shunt current is close to zero. We therefore use ``polarization Hall effect'' as a phenomenological name for the structured transverse-voltage response and treat collective polarization as a possible interpretation.

Figure~\ref{fig:fig1} illustrates proton transfer along a hydrogen bond as a possible source of local dipolar rearrangement in the hydration environment. Interactions among such rearrangements could contribute to a collective, time-dependent polarization. Theories of cooperative dynamics in biological matter, including those developed by Fr\"ohlich and Davydov~\cite{Frohlich1968,Frohlich1988,Davydov1979}, provide motivation for exploring this possibility. 

Within the proposed phenomenological description, the magnetic field modifies the effective landscape governing the collective state variable. The chiral DNA structure and its hydration environment motivate investigation of field-dependent coupling to polarization dynamics.

\subsubsection*{Discrete response levels and oscillatory scales}

The voltage plateaus and inverse-field oscillations provide complementary descriptors of the field-dependent response. In a metastable-state framework, plateaus represent intervals during which the measured voltage remains near a characteristic level, while transitions reflect changes in the stability or occupation of accessible configurations. The approximately regular plateau spacing motivates examination of the structure of the effective dynamical landscape.

The oscillatory measurements introduce a further constraint: a characteristic periodicity in $1/B$. A quantitative model should explain this dependence alongside the voltage thresholds and plateau spacing. The effective density scales derived from the adopted parameterization, approximately $(4.7$--$5.8)\times10^{14}\,\mathrm{m}^{-2}$, offer a comparison between datasets. Assigning these scales to molecular populations or spatial domains requires an explicit physical relation to the measured observables. Studies of hydrogen-bond networks in water~\cite{DonadioGalli2025} provide context for investigating possible structural contributions.

\subsubsection*{Reorganization of the oscillatory response}

Above the threshold in selected field sweeps, dense oscillatory structures evolve into broader, slowly varying lobes. This change, also discussed in earlier measurements~\cite{Pietruszka2025b}, suggests a reorganization of the dominant dynamical timescales or participating modes. Within the polarization hypothesis, coupled local rearrangements could produce a more organized macroscopic response as the effective coupling and relaxation conditions change.

Mode selection in driven systems provides a useful conceptual setting for this interpretation. Comparisons with Fr\"ohlich-type cooperative dynamics and with collective ordering in other physical systems~\cite{Anderson1995,Ketterle1999} motivate questions about energy input, dissipation, and mode occupation. For hydrated DNA, these questions can be formulated in terms of the measured spectral content, phase relationships, and response to the control parameters.

\subsubsection*{Temperature dependence of the electrical response}

Temperature-dependent changes in hydrated-DNA electrical responses have been reported previously~\cite{PietruszkaMarzec2024}. In the present cooling measurements, the longitudinal reorganizations near 20.6 and 12~\degC\ and the emergence of large-amplitude transverse oscillations below approximately 12~\degC\ provide characteristic temperatures for the displayed conditions. Their interpretation should account for changes in conductive and polarization contributions, hydration dynamics, and relaxation times. These observations motivate comparison of spectral amplitudes, fluctuation statistics, and switching behavior across temperature ranges within a common dynamical framework.

The Hilbert-phase analysis of Figure~\ref{fig:temperature_hilbert}
adds a dynamical characterization of the low-temperature
oscillatory regime. Below approximately $12\,^\circ\mathrm{C}$,
the transverse response exhibits sustained phase winding at
fixed magnetic field, with an approximately constant mean rate.
The high linear-fit coefficient ($R^2\approx0.9995$) quantifies
the regularity of the accumulated phase over the analyzed
interval, while variations in the trajectory radius indicate
amplitude modulation. Within the proposed phenomenological
framework, this behavior motivates consideration of a running
oscillatory regime alongside metastable configurations.
The observation at constant field further shows that sustained
phase winding can occur during cooling without a magnetic-field
sweep. 

\subsubsection*{Relationships between electrical and photovoltaic channels}

The longitudinal and transverse signals display related oscillatory structures, with a relative phase near $\pi$ in selected intervals. Tensor-based and parametric representations summarize these relationships and their evolution during field sweeps. Their physical interpretation depends on the conductive and polarization contributions, the circuit response, and the sequential acquisition offset between channels.

The photovoltaic plate, positioned approximately 3~cm from the sample, registered a signal whose phase tracked that of $V_{xy}$ over extended intervals. The longest interval with sliding-window phase agreement $R\geq0.8$ extended from approximately 0.414 to 0.539~T (Figure~\ref{fig:fig11}). A model based on the Cu--solution interface must therefore account for both the transverse voltage at the sample contacts and its phase relationship with the signal recorded at the spatially separated detector.

\subsubsection*{Scope of the control comparisons}

The water control lacks the pronounced structures observed in the hydrated DNA measurements, while DNA samples prepared without added TE buffer retain thresholds and plateau-like levels. The Cu-coated control yields a transverse response of opposite sign. These comparisons establish that the measured response depends on the sample under the conditions tested. The metallic Cu control probes the substrate response, whereas the water control probes the Cu--water interface. The Cu--solution interface in the DNA measurements may respond differently, so electrode polarization and possible Faradaic contributions remain relevant to the interpretation of $V_{xy}$.

\subsubsection*{Connection to earlier measurements}

The present study brings together thresholds, oscillatory structures, and plateau-like voltage levels reported across hydrated-DNA measurements~\cite{Pietruszka2025a,Pietruszka2025b}. Comparisons between these datasets should retain the concentration, temperature, preparation conditions, and acquisition protocol associated with each trace. Recurring characteristic scales motivate a common phenomenological description, while differences between measurements provide constraints on its applicability. This approach preserves the experimental observations and develops their interpretation through explicit links between the measured voltages and the proposed dynamics.

The threshold and staircase reported in Physica B~\cite{Pietruszka2025a} were measured at $21.9\,^\circ\mathrm{C}$, within the temperature range where conductive and polarization contributions may coexist. Subsequent cooling measurements reported in BioSystems~\cite{Pietruszka2026BioSystems} motivate extending the interpretation to include polarization dynamics, without excluding a conductive contribution to the earlier response. The present phenomenological model explores metastable voltage configurations as a possible description of the plateau-like levels.

\subsubsection*{Phase-dependent signal reconstruction}

The numerical superposition in Figure~\ref{fig:fig10} visualizes how the longitudinal and transverse signals combine as an imposed time shift is varied. The alternating regions of enhanced and reduced amplitude reflect the oscillatory content of the recorded signals and provide a useful representation for comparison with their separately extracted phases.

\subsubsection*{Transverse responses and neutral excitations}

Studies of transverse responses associated with neutral excitations provide a conceptual context for exploring polarization-based mechanisms. Relevant examples include phonon-related responses~\cite{Jin2025}, thermal Hall phenomena~\cite{Strohm2005,Grissonnanche2019,ZhangEtAl2023,Xiang2025}, magnon Hall currents~\cite{Onose2010}, and exciton Hall responses~\cite{Mak2018}. These systems illustrate how the nature of the excitations, symmetry, and coupling to external fields enter the description of transverse observables.

For hydrated DNA, the measured observable is an electrical potential difference across the sample. Connecting this voltage to a collective polarization mechanism requires a model of field-dependent dynamics and their coupling to the electrodes. The neutral-excitation literature motivates this investigation, while the thresholds, plateau spacing, and oscillatory scales reported here define the experimental features to be explained.

\subsubsection*{Chirality, geometry, and collective response}

The helical structure of DNA and the organization of its hydration environment offer possible routes for geometry-dependent coupling between magnetic field and polarization dynamics. Within the proposed interpretation, chirality may influence the symmetry and direction of the transverse response. The present measurements motivate this hypothesis, whose quantitative development should specify how molecular geometry enters the effective coupling and how it relates to the macroscopic electrode potentials.

\subsubsection*{Perspectives on proton-mediated polarization}

The observed voltage structures motivate investigation of proton-mediated polarization as a possible source of collective dynamics in hydrated DNA. The hydrogen-bond network provides a physical setting in which local charge rearrangements, hydration structure, and molecular geometry may interact. Studies of coherent energy transfer in photosynthetic complexes~\cite{Engel2007,Panitchayangkoon2010} offer broader context for collective phenomena in biological environments, with distinct observables and mechanisms.

The present findings define concrete targets for a dynamical description: voltage thresholds, metastable response levels, inverse-field oscillatory scales, and temperature-dependent changes in longitudinal and transverse signals. A model connecting these features to conductive and polarization contributions would clarify the physical meaning of the proposed polarization Hall effect and guide further analysis of hydrated soft matter.

\section{Minimal theoretical description}

We introduce a dimensionless collective coordinate $x(t)$ describing the polarization configuration of the hydrated sample. A minimal phenomenological readout relation is
\begin{equation}
V_{xy}(t)=V_{\mathrm{ohm}}(t)+\alpha x(t)+\beta\dot{x}(t),
\label{eq:voltage_readout}
\end{equation}
where $V_{\mathrm{ohm}}$ represents the conductive contribution to the transverse readout, including its dependence on sample and contact geometry. The coefficients $\alpha$ and $\beta$, with units of voltage and voltage times time, respectively, describe the coupling of the polarization configuration and its rate of change to the detected voltage. They may depend on temperature and measurement conditions. Residence near a stable configuration produces a plateau through the $\alpha x$ term, while motion between configurations can generate a transient contribution through $\beta\dot{x}$.

The coordinate $x$ labels successive configurations and is treated as an unwrapped variable. As an illustrative model, we consider overdamped dynamics in a tilted periodic landscape,
\begin{equation}
\gamma\dot{x}
=-\frac{\partial U(x;B,T)}{\partial x}+\xi(t),
\label{eq:configuration_dynamics}
\end{equation}
with
\begin{equation}
U(x;B,T)
=U_0(B,T)\left[1-\cos(2\pi x)\right]-f(B,T)x.
\label{eq:effective_landscape}
\end{equation}
Here $U_0>0$ is an energy scale, $f$ is an effective generalized driving force with units of energy, $\gamma$ is a damping coefficient, and $\xi(t)$ represents fluctuating forces. The field and temperature dependence of $U_0$ and $f$ parameterizes changes in the effective dynamics under the experimental bias and environmental conditions. Local minima provide metastable configurations, while changes in the landscape or fluctuating forces can induce switching between neighboring minima.

For fixed parameters and negligible fluctuations, stationary configurations satisfy
\begin{equation}
\sin(2\pi x)=\frac{f}{2\pi U_0}.
\end{equation}
Stable minima exist for $|f|<2\pi U_0$ and have positive curvature. The condition
\begin{equation}
|f(B,T)|=2\pi U_0(B,T)
\label{eq:depinning_condition}
\end{equation}
marks their disappearance and provides an illustrative depinning criterion. Relating this condition to an experimental threshold requires specification of the field and temperature dependence of the model parameters.

Neighboring minima are separated by $\Delta x=1$. For approximately constant readout coefficients and conductive background, Eq.~\eqref{eq:voltage_readout} therefore gives a characteristic plateau spacing
\begin{equation}
\Delta V_{xy}\approx\alpha.
\label{eq:plateau_spacing}
\end{equation}
Fluctuation-assisted switching between minima can generate discrete changes in voltage, with transient contributions from $\beta\dot{x}$. The periodic landscape illustrates how metastability and approximately regular voltage levels can arise; reproducing the observed staircase during a field sweep requires a specified parameter evolution and switching dynamics.

This model provides explicit relations for testing a configuration-based interpretation of the voltage response. Quantitative application requires parameter estimation and comparison with the recorded traces. Explaining the observed inverse-field periodicity further requires a physical account of the field dependence of the effective landscape.

\section{Conclusion}

Hydrated genomic DNA in a quasi-two-dimensional layer exhibits sharp transverse-voltage thresholds, staircase-like plateaus, and oscillations approximately periodic in inverse magnetic field. In separate cooling measurements, the longitudinal response reorganized near 20.6 and $12.0\,^\circ\mathrm{C}$, while the transverse voltage developed sustained Hilbert-phase winding below approximately $12\,^\circ\mathrm{C}$ at fixed magnetic field, with a mean winding frequency near $0.058~\mathrm{Hz}$. During a field sweep, the voltage recorded by a photovoltaic plate approximately 3~cm from the sample showed extended intervals of phase agreement with $V_{xy}$. The proposed collective-polarization model relates the transverse-voltage plateaus to metastable configurations and gives a depinning criterion and a relation for the plateau spacing.

\section*{Methods}

\subsubsection*{Sample preparation}

Barley genomic DNA (double-stranded, isolated using the CTAB method, modified after Doyle and Doyle~\cite{DoyleDoyle1987}) was dissolved in TE buffer (pH~8, containing Tris--HCl and EDTA). DNA concentrations were 100, 480, 500, and 1000~ng~$\mu$L$^{-1}$ across the measurements reported here. For the measurements designated ``buffer-free,'' the isolated DNA was dissolved directly in distilled water without added TE buffer. These samples still contained DNA-associated counterions and could contain residual salts from the isolation procedure. Counterion identity and residual salt concentrations were not determined for these preparations. The comparison therefore tests whether the observed voltage features persist without added TE. Prepared DNA solutions were stored at $-20~^\circ$C between experiments to minimize degradation.

A 5~$\mu$L droplet was deposited onto a planar four-contact Cu-coated substrate and confined beneath a microscope glass coverslip. The sample footprint varied approximately from $7\times7$ to $10\times10$~mm$^2$, covering the contact pads. Dividing the deposited volume by this area gives a nominal mean layer thickness of approximately 50--100~$\mu$m. The lateral dimensions substantially exceed this thickness, defining the quasi-two-dimensional sample geometry used here. Measurements were performed at ambient relative humidity of approximately 25--35\%. The coverslip confined the deposited liquid over the sample footprint described above. Control measurements were performed on pure water, buffer-free DNA solutions, and metallic Cu-coated substrates.

\subsubsection*{Magnetic-field generation and calibration}

\begin{figure}[t]
  \centering
  \includegraphics[height=0.25\textheight,angle=-90,keepaspectratio]{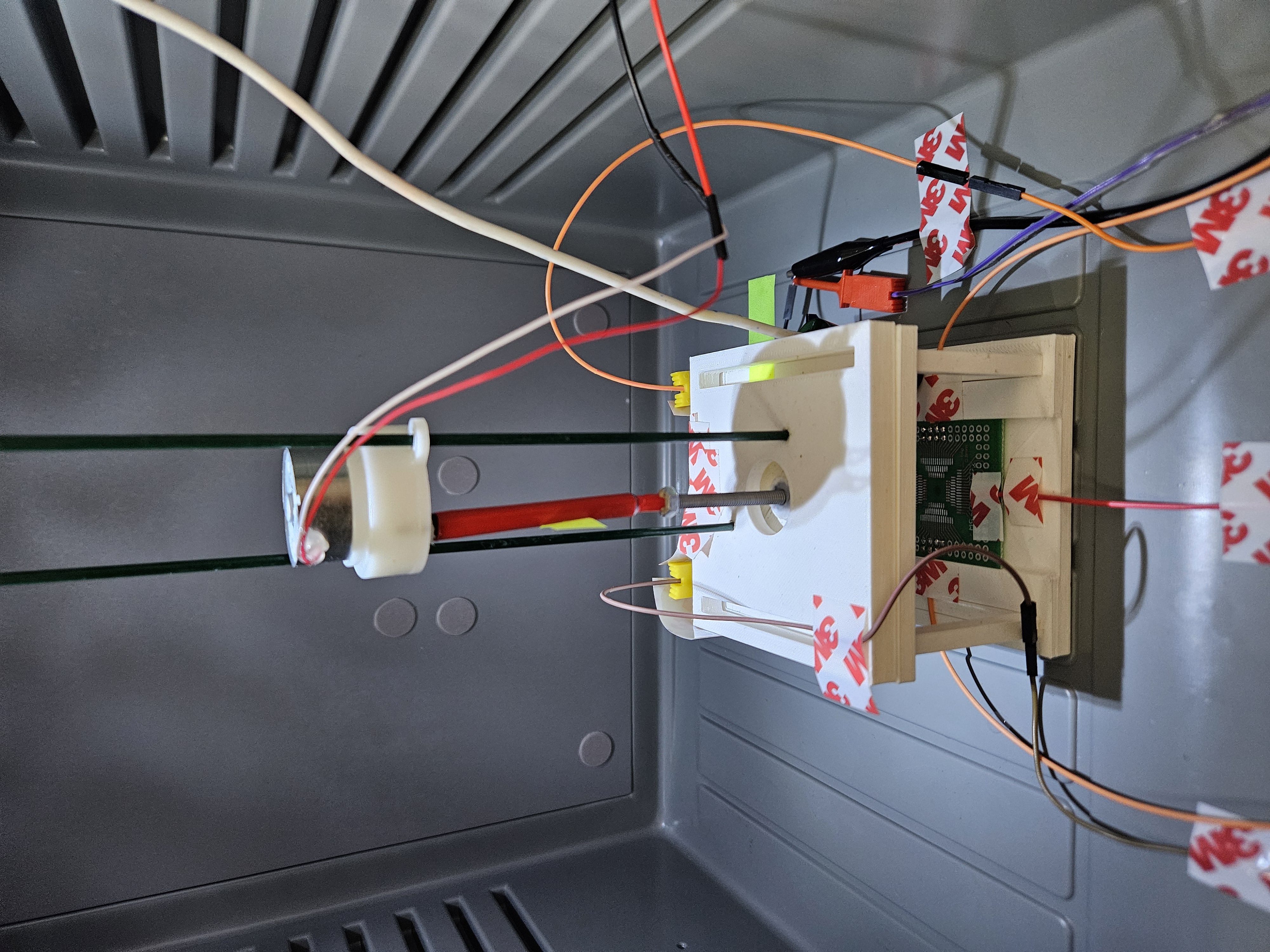}
  \caption{Custom detector for transverse-voltage measurements in hydrated DNA.
A sample of approximately 5~$\mu$L was deposited on a single-sided printed-circuit board (40.5$\times$43~mm) and confined beneath a glass coverslip. The sample footprint varied approximately from $7\times7$ to $10\times10$~mm$^2$, giving a nominal mean liquid-layer thickness of 50--100~$\mu$m from the volume-to-area estimate. Copper contacts provided the electrical connections~\cite{Pietruszka2026BioSystems}. Black contacts supplied the bias, magenta contacts monitored the longitudinal voltage, and yellow contacts recorded the transverse voltage. A perpendicular magnetic field was applied using permanent neodymium magnets. The assembly was placed in a refrigerated, shielded enclosure. Magnet position was varied continuously using the mechanical drive or adjusted manually, as specified for the corresponding experiment.}
  \label{fig:detector}
\end{figure}

Magnetic fields in the range 0.05--0.65~T were generated using NdFeB permanent magnets mounted on a motor-driven linear translation stage that controlled the magnet--sample separation (Figure~\ref{fig:detector}). The system operated in two modes: a motor-sweep mode (0.1--0.5~mT~s$^{-1}$) for high-resolution oscillation and plateau measurements, and a manual mode for threshold measurements to minimize potential electromagnetic interference from the motor. Magnetic-field values were calibrated before each measurement session using a Hall-probe gaussmeter with an accuracy of $\pm1$~mT. All fields were applied perpendicular to the sample plane.

\subsubsection*{Temperature control}

Temperature-dependent experiments were conducted in a refrigerated enclosure equipped with two ice reservoirs to provide passive thermal stabilization. An Extech thermometer was positioned within 5~mm of the sample, and temperature readings were recorded to the nearest $0.1\,^\circ\mathrm{C}$. Typical cooling rates were 0.01--0.03~$^\circ$C~s$^{-1}$. 
The sample footprint was approximately $7\times7$ to $10\times10$~mm$^2$, with a nominal mean liquid-layer thickness of 50--100~$\mu$m. Temperature annotations identify the recorded stages of cooling; the cooling rate may vary during a measurement. Changes in the longitudinal response were observed near 20.6 and 12~$^\circ$C under the conditions presented in Figure~\ref{fig:fig2}. Ambient humidity was monitored during the measurements. Thermal gradients across the sample and contacts are relevant to assessing possible thermoelectric contributions to the voltage response.

\subsubsection*{Electrical measurement geometry}

A 0.1~V DC source was applied through a 1~k$\Omega$ shunt resistor, defining the longitudinal current channel. Signals were recorded using four channels acquired within the same measurement run:
\begin{itemize}
\item CH1: longitudinal voltage \Vxx\ across the 1~k$\Omega$ shunt (yielding $\Ixx=\Vxx/1~\mathrm{k}\Omega$),
\item CH2: longitudinal sample voltage \Vxx\ (sample),
\item CH3: transverse voltage \Vxy\ (sample),
\item CH4: photovoltage (when measured).
\end{itemize}

In the previously reported cooling run at $B=0.5~\mathrm{T}$, the longitudinal current inferred from the $1~\mathrm{k}\Omega$ shunt dropped from approximately $10~\mathrm{nA}$ to nearly zero near $20\,^\circ\mathrm{C}$~\cite{Pietruszka2026BioSystems}. At the applied bias of $0.1~\mathrm{V}$, the electrical power supplied to the sample circuit just before the drop was approximately $1~\mathrm{nW}$.

The magnetic-field sweep experiment shown in Figure~\ref{fig:fig4} was sampled at 740~Hz, yielding approximately $10^5$ data points per trace. The remaining magnetic-field and temperature sweeps were sampled at 1~Hz. The Keithley DMM6500 and the DC power supply were mains-powered, without battery backup or a UPS. The motor driving the magnet-positioning stage was battery-powered. The detector was housed inside the metal enclosure of the refrigerator.

\subsubsection*{Photovoltage measurement}

In selected runs, the photovoltage $V_{\mathrm{PV}}$ was measured at the terminals of a photovoltaic plate positioned approximately 3~cm behind the sample, outside the gap between the magnet and the sample. The plate was connected to CH4 of the same multiplexer card used to record the sample signals on CH1--CH3. The plate had no direct connection to the sample bias circuit. The resulting time series was compared with $V_{xy}$ using the sliding-window phase agreement shown in Figure~\ref{fig:fig11}.

\subsubsection*{Data acquisition and signal processing}

Signals were recorded using a Keithley DMM6500. Longitudinal and transverse channels were acquired sequentially through a multiplexer, with an approximate interchannel offset of 0.25~s in the relevant measurements. Sampling conditions are specified for each dataset.

Low-order polynomial backgrounds were subtracted from selected traces to characterize their oscillatory components. FFT analysis employed a Hann window with amplitude normalization. Analysis in the inverse-field domain yielded characteristic frequencies near 6.3~T for the 5~\degC\ measurement and 1.2~T for the 6~\degC\ measurement. Cross-correlation and Hilbert-phase analysis were used to examine relationships between channels, accounting for their acquisition offset. Spatial coherence lengths require an additional physical model relating temporal phase statistics to distance.

For Figure~\ref{fig:temperature_hilbert}B,C, the transverse
trace was reconstructed from the original figure image by
calibrating its time and voltage axes and extracting the
median position of the red trace in each image column.
The reconstruction was interpolated onto a nominal $1~\mathrm{Hz}$
grid. The power spectral density was calculated over
$920$--$1190~\mathrm{s}$ after mean subtraction, using a Hann
window, and normalized to its maximum. For the complex
trajectory, a third-order Butterworth band-pass filter
($0.035$--$0.075~\mathrm{Hz}$) was applied forward and backward
to avoid phase shifts. The analytic signal was calculated
using the Hilbert transform with 200-sample reflection padding
at each end. Phase was unwrapped, and the linear fit over
$940$--$1160~\mathrm{s}$ was evaluated for the mean-centered
signal without band-pass filtering. Filtering, padding, and
trace-extraction variants gave comparable winding rates.
Reconstruction limits the accuracy of narrow spike amplitudes;
the reported phase-derived values are approximate.

\subsubsection*{Effective response ratios and tensor representation}

The longitudinal and transverse response ratios are defined as
\begin{equation}
R_{xx}=\frac{V_{xx}(\mathrm{sample})}{I_{xx}},
\qquad
R_{xy}=\frac{V_{xy}}{I_{xx}},
\end{equation}
where $I_{xx}$ is obtained from the shunt-voltage measurement. These ratios are undefined at zero current and become sensitive to current uncertainty when $I_{xx}$ approaches zero.

For the tensor-based representation in Figure~\ref{fig:fig7}, the adopted matrix and its inverse are
\begin{equation}
\mathbf{R}=
\begin{pmatrix}
R_{xx} & R_{xy}\\
-R_{xy} & R_{xx}
\end{pmatrix},
\qquad
\boldsymbol{\sigma}_{\mathrm{eff}}=\mathbf{R}^{-1},
\end{equation}
giving
\begin{equation}
\sigma_{xx}=\frac{R_{xx}}{R_{xx}^2+R_{xy}^2},
\qquad
\sigma_{xy}=-\frac{R_{xy}}{R_{xx}^2+R_{xy}^2}.
\end{equation}
This construction assumes equal diagonal and antisymmetric off-diagonal components. The resulting quantities have units of conductance and are used as an effective representation of the measured response, rather than as independently established material conductivities.

\subsubsection*{Extraction of characteristic periodicity scale}

The oscillatory structures are characterized by a frequency $F$ in the inverse-field domain. For the transport-inspired geometric parameterization used in Figure~\ref{fig:fig5}, the derived quantities are
\begin{equation}
A_F=\frac{2\pi e}{\hbar}F,
\qquad
k_F=\sqrt{\frac{A_F}{\pi}}.
\end{equation}
Here $e$ is the elementary charge and $\hbar$ is the reduced Planck constant. These relations specify the numerical conversion used to express the observed periodicity as effective reciprocal-space scales. Their physical interpretation in a polarization framework depends on the proposed coupling mechanism.

Effective density estimates are discussed as quantities derived within the adopted parameterization. Comparisons with plateau-based estimates require the corresponding conversion relations and uncertainty assumptions.

\subsubsection*{Control experiments}

Control comparisons included distilled water, DNA prepared without added TE buffer, and Cu-coated substrates. The displayed water measurement lacks the pronounced structures observed in the DNA traces. Buffer-free DNA retains threshold and plateau-like features, while the Cu comparison gives an opposite-sign transverse response. Forward and reverse field sweeps show approximate agreement in the positions of the principal features. These comparisons characterize the sample dependence of the response under the reported measurement conditions.

\subsubsection*{Response-map construction}

The schematic field--temperature response map was constructed as a qualitative visual summary of selected observations from different DNA concentrations and measurement protocols. Smooth colored regions were used to illustrate the response regimes discussed in the text. Their positions and extents are schematic and were not obtained by fitting or interpolating measured data points. Colors do not encode a measured quantity.

\subsubsection*{Signal reconstruction and measurement contributions}

The numerical reconstruction is defined as
\begin{equation}
S(t,\Delta\phi)=V_{xx}(t)+V_{xy}(t+\tau),
\label{eq:interference}
\end{equation}
where $\tau$ is an imposed time shift. For a narrowband component, $\Delta\phi\approx\omega\tau$. The resulting map visualizes the phase-dependent combination of the recorded signals.

Interpretation of the channel relationships should account for sequential acquisition, preprocessing, circuit coupling, and the shared experimental environment. Possible voltage contributions include conductive and polarization responses, electrode processes, thermal gradients, and electromagnetic pickup. The water, metallic, fixed-field, and photovoltaic comparisons provide complementary information about these contributions within the protocols examined.

\section*{Data availability}

The dataset associated with Figure~\ref{fig:fig4} is available on Zenodo~\cite{PietruszkaZenodo2026}. Additional data and available analysis materials can be obtained from the author upon reasonable request.

\section*{Acknowledgements}

The author acknowledges the use of a Keithley DMM6500 multimeter obtained through a joint study with Marek Marzec reported in \emph{Scientific Reports}. This work was supported by the National Science Centre, Poland (grant no. 2020/37/B/NZ3/03696). The author thanks Katarzyna Konopka for DNA--water samples and Jacek Pietrakowski for technical support.

\section*{References}

\end{document}